\documentclass[prx,twocolumn,tightenlines,aps,epsf,showpacs,superscriptaddress,longbibliography,nofootinbib]{revtex4-2}
\usepackage[pdftex]{graphicx}

\usepackage[normalem]{ulem}
\usepackage{xfrac}
\usepackage{dcolumn}
\usepackage{bm}
\usepackage{epsfig}
\usepackage{latexsym} 
\usepackage{amsmath}
\usepackage{amssymb}
\usepackage{color}
\usepackage{array}
\usepackage{bbm}
\usepackage{color}
\usepackage{array}
\usepackage{cancel}
\usepackage[dvipsnames]{xcolor}

\usepackage[colorlinks=true,allcolors=blue]{hyperref}%

\usepackage{titlesec}
\titleformat{\paragraph}[runin]% runin puts it in the same paragraph
        {\bfseries}% formatting commands to apply to the whole heading
        {}% the label and number
        {0.0em}% space between label/number and subsection title
        {}% formatting commands applied just to subsection title
        [ -- ~]% punctuation or other commands following subsection title
\titlespacing*{\paragraph}{0pt}{4pt}{0pt}

\usepackage[normalem]{ulem} %normalem makes \emph italic rather than underlined

\newcommand{\<}{\langle}
\newcommand{\e}{\varepsilon}
\newcommand{\up}{\uparrow}
\newcommand{\down}{\downarrow}
\renewcommand{\>}{\rangle}
\renewcommand{\(}{\left(}
\renewcommand{\)}{\right)}
\renewcommand{\[}{\left[}
\renewcommand{\]}{\right]}
\renewcommand{\v}[1]{\boldsymbol{#1}} % \v -> vector (bf)

\newcommand{\eps}{\epsilon}

\newcommand{\real}{\Re\text{e}}
\newcommand{\tr}{\text{tr}}

\begin{document}
\title{Holographic  simulation of correlated electrons on a trapped ion quantum processor}
\author{Daoheng Niu}
\email{daoheng@utexas.edu}
\affiliation{Department of Physics, University of Texas at Austin, Austin, TX 78712, USA}

\author{Reza Haghshenas}
\email{haqshena@caltech.edu}
\affiliation{Division of Chemistry and Chemical Engineering, California Institute of Technology, Pasadena, California 91125, USA}

\author{Yuxuan Zhang}
\affiliation{Department of Physics, University of Texas at Austin, Austin, TX 78712, USA}

\author{Michael Foss-Feig}
\affiliation{Quantinuum, 303 S Technology Ct. Broomfield, CO, 80021, USA}

\author{Garnet Kin-Lic Chan}
\affiliation{Division of Chemistry and Chemical Engineering, California Institute of Technology, Pasadena, California 91125, USA}

\author{Andrew C. Potter}
\affiliation{Department of Physics and Astronomy, and Stewart Blusson Quantum Matter Institute,
University of British Columbia, Vancouver, BC, Canada V6T 1Z1}
\begin{abstract}
We develop holographic quantum simulation techniques to prepare correlated electronic ground states in quantum matrix product state (qMPS) form, using far fewer qubits than the number of orbitals represented. 
Our approach starts with a holographic technique to prepare a compressed approximation to electronic mean-field ground-states, known as fermionic Gaussian matrix product states (GMPS), with a polynomial reduction in qubit- and (in select cases gate-) resources compared to existing techniques. Correlations are then introduced by augmenting the GMPS circuits in a variational technique which we denote GMPS+X. 
We demonstrate this approach on Quantinuum's System Model H1 trapped-ion quantum processor for 1$d$ models of correlated metal and Mott insulating states.
Focusing on the $1d$ Fermi-Hubbard chain as a benchmark, we show that GMPS+X methods faithfully capture the physics of correlated electron states, including Mott insulators and correlated Luttinger liquid metals, using considerably fewer parameters than problem-agnostic variational circuits.
\end{abstract}
\maketitle

%\paragraph{Introduction}
As quantum computers have begun to achieve the scale and reliability required to surpass classical computations on certain theoretically-contrived tasks such as random quantum circuit sampling~\cite{arute2019quantum,zhong2020quantum,aaronson2011computational}, it is natural to ask 
how best to apply their computational abilities 
%when their computational power could be applied 
to problems of technological and scientific interest. %First-principles 
%The simulation of electronic, magnetic, optical, and chemical properties of correlated molecules and materials are some of the most 
The quantum simulation of molecules and materials is a promising target application, where there are theoretical grounds to expect exponential quantum computational advantage~\cite{bauer2020quantum} in certain types of quantum dynamics, with prospective applications to non-equilibrium electron transport, quantum reactive scattering, and the dynamics of strongly-coupled field theories.
An important prerequisite to computing dynamics in physical applications is to first prepare a good approximation to the ground- or thermal-state of a correlated electron system. 
In variational approaches, a key goal is to use physically-inspired circuit ansatzes to reduce the number of variational parameters and simplify the optimization landscape.
Matrix-product states (MPS)~\cite{schollwock2011density} provide a compact parameterization of many physically-important quantum states since the memory and complexity of MPS calculations are controlled by the extent of spatial correlations and entanglement, encoded by the matrix size (``bond-dimension'') $\chi$.
A growing body of work~\cite{barratt2021parallel,foss2021holographic,lin2021real,foss2021entanglement,chertkov2021holographic,slattery2021quantum,maccormack2021simulating,zhang2022qubit} has begun to extend the efficient data compression afforded by classical MPS techniques to the quantum domain, using quantum memories with $\sim \log_2\chi$ qubits to represent the bond-space of an MPS, and quantum circuits interleaved with partial-measurement to implement its tensors. By exploiting mid-circuit measurements and qubit reuse (MCMR)~\cite{foss2021holographic}, a quantum MPS (qMPS) simulation of a $d$-dimensional systems can be performed with effectively $(d-1)$ dimensions' worth of qubits, earning the moniker ``holographic simulation"~\cite{kim2017holographic,foss2021holographic}.  
\begin{figure*}[t] 
    \centering
    \includegraphics[width=\textwidth]{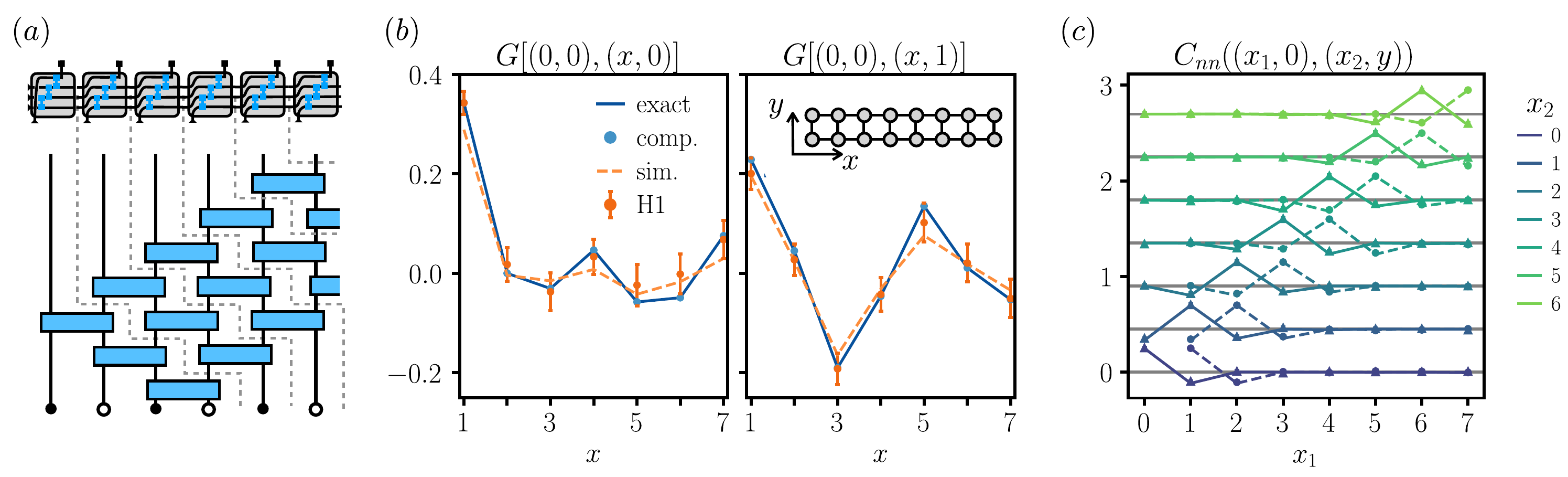}
    \caption{ {\bf Compressing Gaussian Fermion States as qMPS -- } 
    \label{fig:free} 
    (a) Holographic qMPS implementation (top) of the GMPS circuit (bottom) for approximately preparing compressed Gaussian fermion states.  (b-c) Experimental implementation of holographic GMPS algorithm for a spinless two-leg ladder at half-filling. (b) (Real-part of) Green's function, $\real~ G\[\v{r}=(0,0),\v{r}'=(x,y)\]$ with experimental data (orange dots with $1\sigma$ error bars from $1000$ measurement shots per point), noisy circuit simulations with one-qubit and two-qubit gate depolarizing parameters $p_{1q}=10^{-3}$ and $p_{2q}=5\times 10^{-3}$ respectively (dashed orange line), and exact values (solid blue line)). (c) Connected part of density-density correlators
    $C_{nn}(\v{r},\v{r}')$ with $\v{r} = (x_1,0)$ and $\v{r}' = (x_2,y)$, with $y=0$ data shown as solid-lines (theory) and triangles (experiment), and $y=1$ data shown as dashed-lines (theory) and circles (experiment) respectively. Each point represents $5600$ measurement shots. Statistical error bars in (c) are included, but smaller than plot symbols.}
        \label{fig:GMPS}
\end{figure*}
Early demonstrations of holographic qMPS~\cite{barratt2021parallel,foss2021holographic,foss2021entanglement} and their higher-dimensional quantum tensor network (qTNS) generalizations~\cite{slattery2021quantum,maccormack2021simulating} have focused on simple spin models.
However, more realistic molecular and material models contain fermionic electrons, which commonly exhibit quite entangled (e.g. metallic) ground-states even for weak interaction strengths, and whose Hamiltonians have more complex qubit representations.

Because many correlated electron states are adiabatically connected to Gaussian (e.g. mean-field) states, we first describe how to prepare holographically Gaussian MPS (GMPS) as a qMPS.
We start from a classical construction of Fishman and White~\cite{fishman2015compression} that naturally compresses Gaussian fermionic states in GMPS form. A key aspect of this compression is that,
unlike other mean-field preparation techniques, it exploits the near-area-law nature of the ground-state to parametrically reduce qubit and gate resources.
We present numerical evidence that this holographic GMPS method gives a polynomial reduction in the number of qubits (and in certain cases, gates) 
compared to standard quantum algorithms.
Next, using the $1d$ Fermi-Hubbard chain as a benchmark, we show that augmenting the mean-field GMPS state
with shallow circuits (which we refer to as GMPS+X) enables more accurate variational preparation of correlated electron ground-states with far-fewer parameters than problem-agnostic ansatzes, a key component for scaling
%raises the possibility of reducing a key obstacle to scaling 
up qMPS methods to solve larger and more classically-challenging models. % complicated, classically intractable models.
%
%A related point is that, while one could in principle simple work directly in the basis of mean-field orbitals (as is common in quantum chemistry applications), this would obscure the spatially locality of the interaction Hamiltonian, and spoil the ability to effectively compress the state into qMPS form.
We implement the GMPS and GMPS+X methods on Quantinuum's system model H1 trapped-ion quantum processor, demonstrating that the resource reductions enable the faithful simulation of systems with up to 16 orbitals
using minimal error mitigation.
Finally, we explore extending the GMPS compression to $2d$, where we show that the required qubit resources continue to be asymptotically optimal, and discuss the potential advantages of this method for simulating correlated topological phases.
%alarge systems with a small number of qubits, allowing accurate simulation with minimal error-mitigation. 
%
%To our knowledge, the GMPS+U implementation on a six-site Hubbard chain represents the largest-scale simulation of a correlated electron ground-state on a programmable quantum device.\red{\bf Is this true?}

%Finally, we argue that as a corollary, these techniques that a large class of physically relevant states with exponentially large (in $L$) bond-dimension MPS representation, can be efficiently holographically represented as a qMPS with polynomial scaling of qubit- and gate- resources, and establishes a polynomial reduction in qubit resources compared for holographic simulation of gapped states without fractionalized excitations (including correlated trivial and topological band insulators and superconductors) compared to standard quantum simulation methods.

\section{Quantum Matrix Product States (qMPS)}
Here, we briefly recap the holographic simulation with qMPS. Interested readers may find a more detailed exposition in Ref.~\cite{foss2021holographic}.
Holographic simulation with qMPS is based on the matrix-product state (MPS) representation:
\begin{align}
	|\Psi\> = \sum_{n_1\dots n_L} \ell^T A^{n_1}A^{n_2}\dots |n_1n_2\dots n_L\>
\end{align}
where $n_x\in \{1,2,\dots d=2^{N_p}\}$ label the orbital occupation numbers of the $N_p$ different physical spin- or orbital- ``flavors" at position $x$, $A^{n_x}$ is a $\chi \times \chi$ matrix for each $n_x$ label, and $\ell$ is a $\chi$-dimensional vector that determines the left boundary-conditions. $2d$ and $3d$ systems can also be treated in this framework by treating the system as a $1d$ stack of $(d-1)$-dimensional cross-sections. In this case, $\chi$ must scale exponentially in the cross-sectional area for area-law states.

Properties of any MPS in right-canonical form (RCF)~\cite{perez2006matrix} can be measured on a quantum computer by implementing its transfer-matrix as a quantum channel~\cite{gyongyosi2012properties} acting on $N_p=\log_2 d$ ``physical" qubits and $N_b=\log_2\chi$ bond qubits~\cite{foss2021holographic}. The bond-vector $\ell$ is prepared by acting on the bond-qubits plus optional ancilla with a unitary $U_\ell$. Each tensor $A$ is then embedded into a  larger block unitary operator $U_A$ acting on a reference initial state, $|0\>$, of the physical qubits: $A^n_{ij} = \<n|_p\otimes\<i|_b U_A |0\>_p\otimes |j\>_b$ where subscripts $p$ and $b$ respectively denote physical and bond qubits. Then, the physical qubits can be measured in any desired basis (without measuring the bond qubits). The process is then repeated for each site in sequence from left to right until the measurement is completed.
As for any quantum algorithm, repeated statistical sampling of these measurements must be used to estimate the expectation values of observables.
In this way, one can measure any product operator of the form $\prod_{x=1}^L \mathcal{O}_x$, which forms a complete basis for general observables. 
To summarize, the qMPS procedure for sampling an observable of the form $\<\psi|\prod_{x=1}^L \mathcal{O}_x|\psi\>$ is: 

\begin{enumerate}
\item Prepare the bond qubits in a state corresponding to the left boundary vector $\ell$.
\item Reset the physical qubit for site $[x=0]$ in a fixed reference state $|0\>$.
\item Perform a quantum circuit representing $U_A$ at site $[x]$, entangling the physical and bond qubits. \item Measure the physical qubit in the eigenbasis of $O_x$ and weight the measurement outcome by the corresponding eigenvalue of that observable. The bond-qubit register now corresponds to bond connecting sites $x$ and $x+1$.
\item Repeat steps 1-4 for $x=1\dots L$, and discard the bond-qubits~\footnote{In the last step, there is no reason to continue the chain beyond site $x=L$ where $L$ is the rightmost site where the observable has support: since we are preparing a state in RCF, the tensor contractions without operator insertions for $x>L$ simply multiply the $x\leq L$ network from the right by the identity vector, which corresponds to tracing out or discarding the bond qubits.}.
\end{enumerate}

Since the physical qubits for site $x$ are reset and reused as physical qubits for site $x+1$, this qMPS procedure saves the total number of qubits to be used and enables a small quantum processor to achieve quantum simulation tasks with a bigger size than its available number of qubits.
Moreover, the entanglement spectrum of the bond-qubits in between sites $x$ and $x+1$ coincides with the bipartite entanglement spectrum of the physical MPS at that entanglement cut, further enabling measurement of non-local entanglement observables, as recently demonstrated experimentally for near-critical spin chains~\cite{foss2021holographic}.

\section{Models and Observables}
We focus on quasi-1d Fermi-Hubbard (FH) models, which we write in the form:
\begin{align}
\label{eq:FH}
H_\text{FH} = -t\sum_{\sigma,\<\v{r},\v{r}'\>}c^\dagger_{\v{r},\sigma}c^{\vphantom\dagger}_{\v{r}',\sigma} +\frac{U}{2}\sum_{\v{r}} n_{\v{r}}\(n_{\v{r}}-1\)-\mu N
\end{align}
where $c^\dagger_{\v{r},\sigma}$ creates an electron at site $\v{r}=(x,y)$ with $z$-component of spin $\sigma\in \{\up,\down\}$, $1\leq x,y\leq L_{x,y}$, $n_{\v{r}}=\sum_\sigma c^\dagger_{\v{r}\sigma}c^{\vphantom\dagger}_{\v{r}\sigma}$, $N=\sum_{\v{r}}n_{\v{r}}$, $t$ is the hopping strength, $U$ is the onsite Hubbard interaction, and $\mu$ is the chemical potential. We measure three types of observables, single-particle equal-time Green's functions (a.k.a. one-electron density matrices), $G$, and connected density-density and spin-spin correlators, $C_{nn}$ and $C_{SS}$:
\begin{align}
G_{\v{r},\sigma;\v{r}',\sigma'} &= \<c^\dagger_{\v{r},\sigma}c^{\vphantom\dagger}_{\v{r}',\sigma'}\> 
\nonumber\\
C_{nn}(\v{r},\v{r}') &= \<n_{\v{r}}n_{\v{r}'}\>-\<n_{\v{r}}\>\<n_{\v{r}'}\>
\nonumber\\
C_{SS}(\v{r},\v{r}') & = \<S^z_{\v{r}}S^z_{\v{r}'}\>-\<S^z_{\v{r}}\>\<S^z_{\v{r}'}\>
\end{align}
where $S^z_{\v{r}} = \frac12 \sum_{\sigma,\sigma'} c^\dagger_{\v{r},\sigma}\sigma^z_{\sigma\sigma'}c^{\vphantom\dagger}_{\v{r},\sigma'}$.

To simulate fermionic systems, one needs to encode the physical fermionic orbital creation and annihilation operators into bosonic qubit degrees of freedom. A variety of encodings are available. Throughout this work, we adopt the Jordan-Wigner (JW) encoding, which is natural for quasi-$1d$ settings, with the convention that orbitals are ordered first by spin $\{\up,\down\}$, then by ascending $y$-position, and finally by ascending $x$-position. We remark that the holographic representation in principle enables (Appendix~\ref{appendix:noJW}) one to reduce the maximal length of JW strings that one needs to measure in variational calculations from $\sim L^d\rightarrow L^{d-1}$, reducing the impact of measurement errors in computing long-distance correlation functions.
%The chief complication introduced by this encoding is that measurements of two-point fermion-correlations require measuring long products of $\sigma^z$ Pauli operators. For very long strings or large measurement error rates this generically results in asymptotic exponential decay of measured correlators. While we do not face this difficulty on the scale of experiment performed in this paper, in Appendix~\ref{} we present a trick to avoid this difficulty by pulling the extended JW string back into the bond space where the $1d$ string reduces to its $0d$ end-points -- highlighting another potential advantage of holographic simulation techniques for fermion encoding.

\section{Compressing Gaussian States as qMPS}
We begin by briefly reviewing the classical MPS algorithm to construct an MPS representation of Gaussian fermion states, and explain how to recast the resulting Gaussian MPS (GMPS) as a qMPS.

\subsection{Compressed Gaussian MPS (GMPS)}
The ground-state of a non-interacting fermion Hamiltonian with $N_{\rm o}$ orbitals: $H=\sum_{i,j=1}^{N_{\rm o}} c^\dagger_i h_{ij} c^{\vphantom\dagger}_j$ is fully characterized by its $N_{\rm o}\times N_{\rm o}$ single-particle Green's function: $G_{ij} = \<c^\dagger_i c^{\vphantom\dagger}_j\>$ (generalizations to non-number conserving Hamiltonians are discussed in Appendix~\ref{appendix:non-conserving}), which has highly-degenerate eigenvalues of either 0 or 1 and whose eigenvectors correspond to unoccupied or occupied orbitals respectively. The Green's function is preserved by any unitary transformation acting separately on the (un)occupied subspaces. The compression scheme of~\cite{fishman2015compression} exploits this freedom by progressively disentangling well-localized degrees of freedom in blocks of $B$ adjacent sites, where $B$ is chosen to be sufficiently large to achieve target infidelity, $\eps$. Starting with the upper-left $B\times B$ block of $G$, one finds the eigenvector of this sub-block whose eigenvalue is closest to either $0$ or $1$ and performs a series of $2\times2$ (single-particle) unitary rotations that move this eigenvector to the first site of the block. The procedure is iterated for the remaining $(N_{\rm o}-1)\times(N_{\rm o}-1)$ sites until the Green's function is approximately diagonalized.

The composition of all the basis rotations in the above procedure produces an $N_{\rm o}\times N_{\rm o}$ unitary, $u^\dagger=(\prod_{\alpha=1}^{(B-1)(N_o - \frac{B}{2})} u_\alpha)^\dagger$, consisting of a ladder of $2\times 2$ single-particle unitaries labeled by ordered index $\alpha$, which approximately diagonalizes the Green's function. Alternatively, read in reverse, the inverse transformation $u$ approximately converts a product state of (un)occupied sites into the entangled ground-state of $h$. These single-particle (size $n\times n$) operations can be converted into a circuit acting on the many-particle Hilbert space (of size $2^n$) by replacing each $2\times2$ unitary, $u_\alpha$, by an equivalent two-qubit gate: 
$U_\alpha = \exp\[\sum_{ij} c^\dagger_i (\log u)_{ij} c^{\vphantom\dagger}_{j}\] = \exp\[\sum_{ij} \sigma^+_i (\log u)_{ij} \sigma^-_{j}\]$, where in the second line we have used the fact that the rotation always occurs between neighboring sites and therefore does not involve a Jordan Wigner string.

Crucially, the resulting ladder-circuit $U=\prod_\alpha U_\alpha$ can be interpreted as a qMPS with bond dimension $\chi=2^B$ by chopping it into diagonal causal slices (see Fig.~\ref{fig:GMPS}), and interpreting the qubit lines entering the bottom of the slice as physical qubits and those entering the side as bond-qubits. We will refer to the resulting MPS as a Gaussian MPS (GMPS) to distinguish it from generic non-Gaussian (q)MPS of the same bond dimension.

Whereas an arbitrary Gaussian state can be prepared by a ladder circuit acting on $N_{\rm o}$ qubits with $O(N_{\rm o}^2)$ two-qubit gates (see for example~\cite{kivlichan2018quantum, arute2020hartree}), the compressed GMPS \emph{ground-state} requires $O(N_{\rm o}B)$ two-qubit gates acting on $O(B)$ qubits (if implemented holographically). The efficiency of this compression therefore depends on the block size $B$ required to accurately approximate the state in question. Empirical numerical evidence and entanglement-based arguments indicate that GMPS for ground-states of local Hamiltonians in $1d$ systems of length $L$ and for target error threshold $\epsilon = 1-\frac{1}{L}\sum_{i,j}|G_{ij}^{(\text{GMPS})}-G_{ij}|$ requires block size (equivalently number of qubits) $B\sim \log \eps^{-1}$ for a gapped state or $B\sim \log L\log \eps^{-1}$ for a gapless metallic state. In Section~\ref{app:2dgmps} below, we extend these results to $2d$-dimensional systems, and find that generically $B$ scales with the bipartite entanglement entropy $S(L)$: 
\begin{align}
B\sim S(L) \sim \begin{cases}
 L\log \eps^{-1} & \text{gapped} \\
  L\log L\log \eps^{-1} & \text{Fermi-surface} 
  \end{cases},
\end{align}
\footnote{
    We note that these asymptotic expressions are asymptotic in $L$ and assume non-zero density of particles. For finite-size molecules there may be individual orbitals that add extra overhead.
}.
We conjecture that similar scalings with $L\rightarrow L^2$ hold in $3d$ (e.g. this follows straightforwardly from the $1d$ results for translation-invariant systems).
 This result holds even for topologically non-trivial Chern band insulators that have an obstruction to forming a fully localized Wannier-basis. 
Combined with holographic simulation methods using mid-circuit measurement and reuse, this method dramatically reduces the number of qubits required ($L^{d-1}B$ vs. $L^d$) to implement the GMPS on a quantum computer.

\subsection{Trapped-ion GMPS implementation}
To demonstrate the feasibility of this approach for near-term hardware, we prepare an 
%highly-
entangled metallic ground-state of a spinless non-interacting two-leg ladder (2ll) described by Eq.~\ref{eq:FH} without spin, and with $L_x=8,L_y=2,U=0$ and $N=8$ electrons (half-filling). 
This system has the same number of degrees of freedom of a spinful $L_x=8$ FH chain that we will be ultimately interested in, but avoids the trivial decoupling of spin-species that arises in mean-field ground-states of the FH chain.
The qMPS representation of the compressed GMPS is implemented on Quantinuum's System Model H1 trapped-ion quantum computer utilizing $6$ trapped-ion qubits corresponding to block size $B=2\times 3$, sufficient to reduce the theoretical compression error below $1\%$.
% %
% \red{The implementation uses 130 controlled-Z gates in total for the system of 16 electrons, which is a similar two-qubit count compared to Google's circuit for a same size system it would need 128 $\sqrt{\mathrm{iSWAP}}$ gates. Asymptotically, using GMPS compression need less two-qubit gates compared to Google's circuit.}  \yz{not very clear what we are trying to express here?} \blue{\it DN: (two sentences above change to:) The implementation uses 130 controlled-Z gates in total and in general it needs $(B-1)(2N-B)$ controlled-Z gates for $N$ sites and block size $B$. In Google's paper~\cite{arute2020hartree} where 72 $\sqrt{\mathrm{iSWAP}}$ gates are used to simulate $H_{12}$ chain using, in general $2(\frac{N}{2})^2$ $\sqrt{\mathrm{iSWAP}}$ gates are needed for $N$ sites. So for a same system size as the 2ll we simulated, Google's algorithm needs 128 $\sqrt{\mathrm{iSWAP}}$ gates. In comparison, the GMPS compression algorithm uses about the same amount of two-qubit gates as Google's algorithm for the system size we simulated but asymptotically needs less two-qubit gates compared to Google's algorithm.}
%
Using only a simple error mitigation scheme based on post-selecting data with the correct total electron number (see Appendix~\ref{app:em} for details), we achieve close to a quantitative agreement (within statistical error bars) between the experimental correlation functions and their theoretical values (see Fig.~\ref{fig:GMPS}).

\section{Correlated electron models}
Since non-interacting fermion systems permit efficient classical simulation, the GMPS technique is not directly useful on its own. However, holographic qMPS approximations to mean-field states can be helpful starting points for approximating correlated ground states either by i) adiabatic evolution to a correlated system in the same phase as the mean-field state (using efficient holographic time-evolution methods~\cite{foss2021holographic}), or ii) reducing the complexity of variational state preparation by providing a good initial guess. Here, we focus on the variational approach ii) and show that relatively simple variational-circuit extensions of the GMPS circuit, which we refer to as GMPS+X methods, provide good approximations to interacting fermion ground-states of a spinful Fermi-Hubbard (FH) chain (Eq.~\ref{eq:FH} with $L_y=1$). 
Since this model can be exactly solved by Bethe-Ansatz methods, it provides a convenient, high-precision benchmark of these techniques.
We implement two different GMPS+X ansatzes, which we will refer to as GMPS+J and GMPS+U respectively, and compare their performance against a problem-agnostic ansatz where the qMPS tensors are generated by a brickwork circuit of general (number-conserving) two-qubit gates.
In each of the GMPS+X approaches, we first construct a GMPS circuit corresponding to the Hartree-Fock (HF) ground-state. At half-filling ($\nu=\sfrac12$, one electron per site), the HF ground-state has antiferromagnetic (AFM) order with  order-parameter: $\mathcal{N} = \sum_j(-1)^j\<S^z_j\> \neq 0$. 
The long-range AFM order is, of course, an artifact of the HF-approximation, and the true ground-state has only power-law decaying AFM correlations due to strong quantum fluctuations.
We also consider one-third-filling ($\nu=\sfrac13$, two electrons per three sites), where we use a HF ground-state solution that is a non-magnetic Fermi-gas, and the true ground-state is a correlated Luttinger-liquid with spin-charge separation.

\begin{figure*}[t] 
    \centering
    \includegraphics[width=0.8\textwidth]{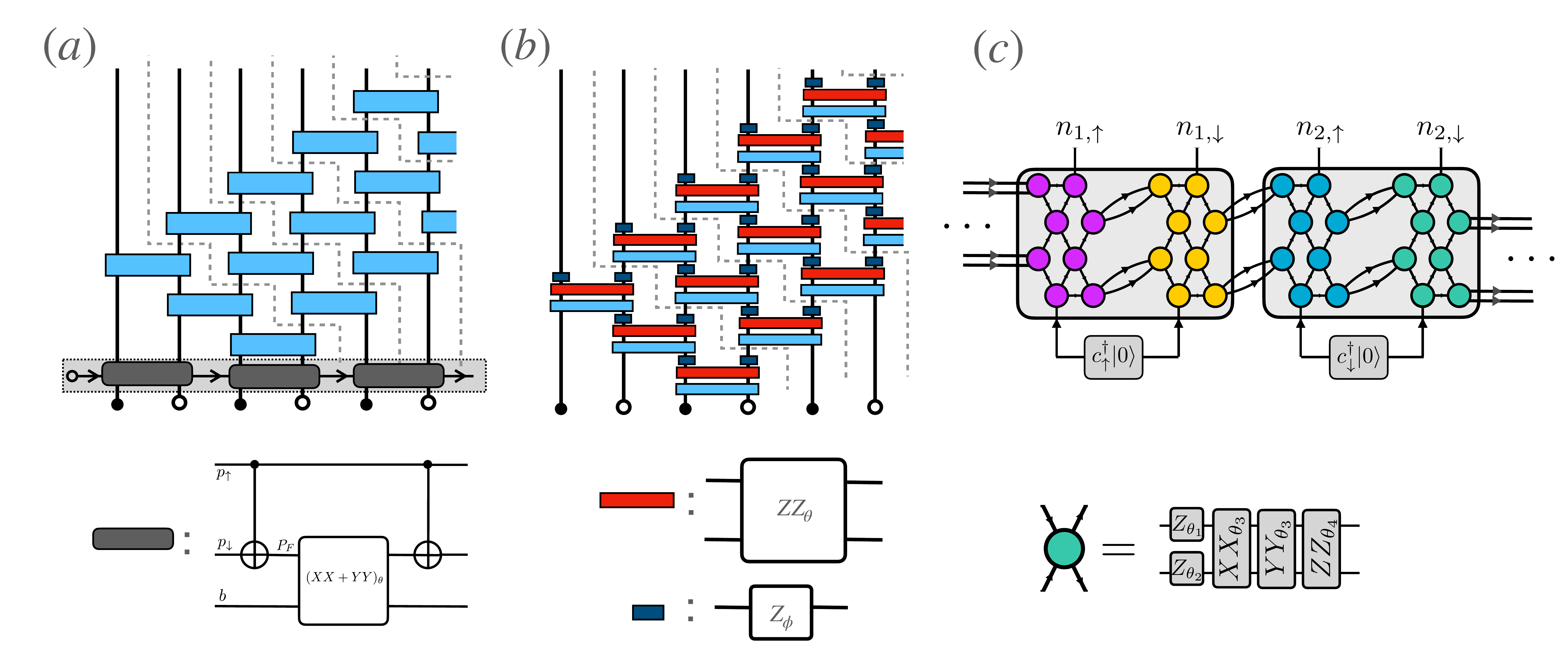}
    \caption{{\bf Variational circuit architectures for correlated electron problems --} The GMPS+X approaches augment the GMPS circuit preparing the Hartree-Fock ground-state with non-Gaussian gates that build in correlations either by  (a) GMPS+J: introducing an extra qMPS layer with an extra bond-qubit (gray dashed box) or by (b) GMPS+U: generalizing the GMPS gates (blue boxes) to include non-Gaussian operations, $ZZ_\theta = e^{-\frac{i}{2}\theta Z\otimes Z}$ and $Z_\phi = e^{-\frac{i}{2} \phi Z}$. Here we draw the GMPS+X circuits only for block size $B=3$ (the implemented circuits have twice this block size, $B=6$ to include spin). (c) A problem-agnostic brick qMPS ansatz consisting of a brickwork of general number-conserving two-qubit gates.
    Here, for any Hermitian operator $O$, $O_\theta$ denotes a gate corresponding to unitary $u[O_\theta] = e^{-i\theta O/2}$.
    \label{fig:vqe}
        }
\end{figure*}
\begin{figure*}[t] 
    \centering
    \includegraphics[width=1.0\textwidth]{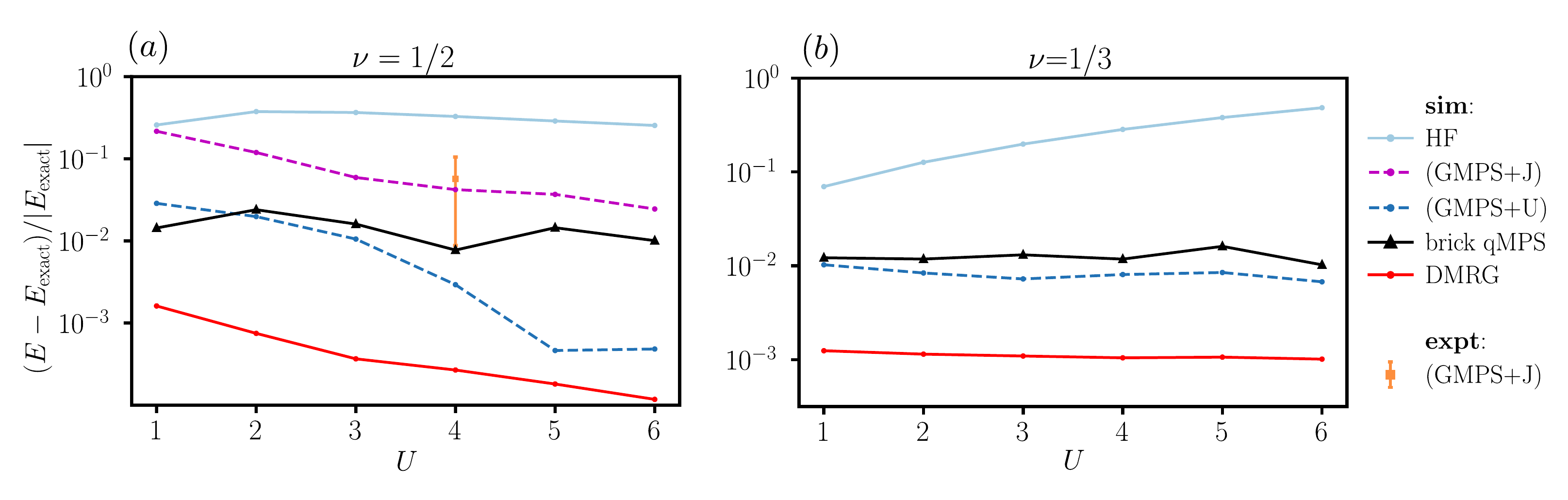}
    \vspace{-0.3in}
    \caption{ {\bf Comparison of variational approaches to the Fermi-Hubbard chain --} \label{fig:GMPS_U} 
    Relative error in energy of (a) half-filling (b) (\sfrac13)-filling Fermi Hubbard chain for various variational approaches compared to the exact ground-state of the corresponding Fermi-Hubbard chain. %
    Variational approaches include two types of GMPS+X circuits: $\rm GMPS+J$ and $\rm GMPS+U$ that augment the GMPS with $B=6$ for the Hartree-Fock ground-state with additional variational circuitry, and a problem-agnostic qMPS ansatz with two bond qubits and four layers of brick circuit (brick qMPS)
    (see Appendix~\ref{appendix:design} for details of circuit ansatzes).
    The GMPS+U simulations were performed for an $L=18$ site chain, whereas the brick qMPS and DMRG calculations were performed for an infinite chain.
    For comparison to classical methods, we include the mean-field (MF) solution, and DMRG results with bond-dimension $2^5$ that provide a lower-bound on the achievable energy with $n_b=5$ bond-qubits ($n_b = B - 1$ in general) used in the GMPS+U approach. The number of variational parameters per (spinful) site are respectively: 1 (GMPS+J), 30 (GMPS+U), 80 (brick qMPS), and $1024$ (DMRG with $\chi=2^5$). 
        }
\end{figure*}
\subsection{Variational ansatzes}
Here we introduce and briefly describe the different variational approaches considered (see Fig.~\ref{fig:vqe}), and compare their performance through numerical simulations. Additional details on the construction and implementation of circuits for each ansatz can be found in Appendix~\ref{app:vqe}.

\paragraph{GMPS+J ansatz} The GMPS+J ansatz is specific to half-filling ($\nu=\sfrac12$), where the charge degrees of freedom are gapped, and the low-energy fluctuations of the FH chain are approximately described by a Heisenberg spin-\sfrac12 chain. 
Roughly-speaking, we can think of the GMPS circuit as transforming from the sites of a spin-\sfrac12 chain to the Wannier orbitals for the Mott insulating FH-chain.
With this picture in mind, we can build in spin-correlations by adding variational layers \emph{before} the GMPS circuit, such that that the GMPS circuit produces an entangled state of the Wannier-orbital spins, rather than a simple Ne\'el product state (Fig.~\ref{fig:vqe}a).
Specifically, we choose a single variational circuit layer that is equivalent (up to a basis change) to that used in \cite{foss2021holographic} to approximate the Heisenberg spin-chain ground-state as a qMPS using a single bond-qubit (see Appendix \ref{app:vqe} for details).
We will see that this ansatz performs best at half-filling and large-$U$, where the FH chain can be well-approximated by a spin-chain.
The chief advantage of this ansatz is that it is very compact, requiring only a single variational parameter per site.
\paragraph{GMPS+U ansatz} 
The GMPS+U ansatz (Fig.~\ref{fig:vqe}b) simply augments each of the Gaussian fermion gates in the GMPS circuit with a non-Gaussian gate $e^{-\frac{i}{2}\(\theta_i Z\otimes Z+\alpha_iZ\otimes 1 +\beta_i 1\otimes Z\)}$ with variational parameters $\{\theta_i,\alpha_i,\beta_i\}$ chosen independently for each GMPS-gate, $i$. These non-Gaussian operations make the GMPS+U gates into a general number-conserving gate (with some parameters fixed by the GMPS representation of the HF state). We will see that this ansatz is more flexible than the GMPS+J method, and can achieve reasonable results over a broad range of fillings and interaction strengths, albeit at the cost of introducing additional variational parameters $n_\text{var}=3N_S(B-1)$, where  $N_S=2$ denotes the number of spin components. We will show that this also implies that the representational power of the GMPS+U ansatz increases with $B$, and will give evidence that this enables the ansatz to be scaled to achieved arbitrary desired variational accuracy (Fig.~\ref{fig:ErrvsB}).

\paragraph{Brick qMPS ansatz}
Finally, we compare the $\rm GMPS+U$ approaches to a problem-agnostic qMPS whose tensors are generated by a brickwork circuit (see Fig.~\ref{fig:vqe}c) of arbitrary (charge-conserving) two-qubit gates.
In this approach, the circuit parameters are not constrained except by symmetry. Specifically, we enforce charge conservation by demanding the circuits commute with the total $S^z$ of the physical and bond-qubits, resulting in $5$ variational parameters per gate\footnote{A general $SU(4)$ unitary has $15$ variational parameters, this can be reduced to $5$ by enforcing symmetry and exploiting the gauge-structure of the matrix product states}.

\subsection{Comparison of variational approaches}
\begin{figure}[t] 
    \centering
    \includegraphics[width=0.5\textwidth]{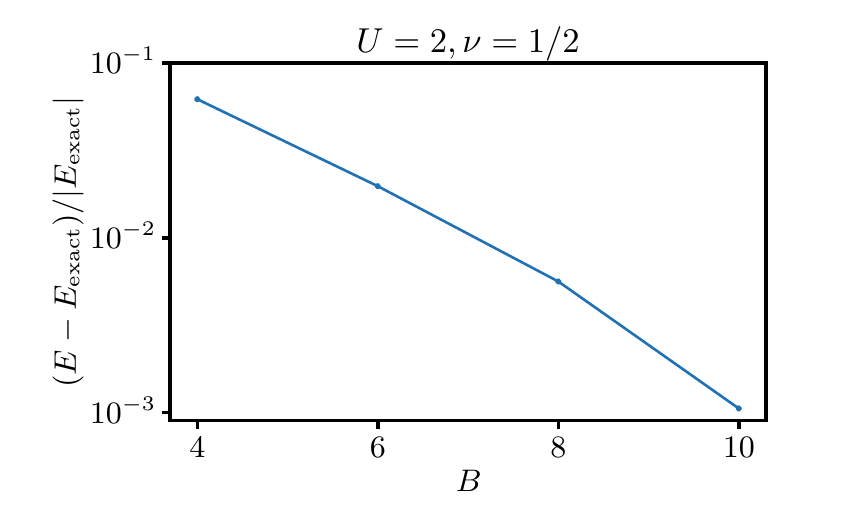}
    \caption{ {\bf Scaling of the GMPS+U method when changing block size $B$ --}
    \label{fig:ErrvsB}
    Variational energy of GMPS+U ansatz for the Fermi-Hubbard chain at half-filling with $U=2$ converges rapidly (approximately exponentially over the range of parameters explored) with block-size $B$, which suggests that this method can be used to achieve arbitrary desired accuracy.  
        }
\end{figure}

Fig.~\ref{fig:vqe} shows numerical results for the relative error $\eps=(E-E_\text{exact})/|E_\text{exact}|$ between the variational energy $E$, and the exact ground-state energy $E_\text{exact}$, for different variational qMPS ansatzes at fillings $\nu=\sfrac12,\sfrac13$ and various interaction strengths $1\leq U\leq 6$. These are additionally compared to the HF approximation and a classical DMRG calculation with bond dimension $\chi=2^5$, equivalent to the bond dimension for the $n_b=5$ bond qubits needed for the GMPS+X approaches. Since DMRG effectively converges to near-optimal results in these type of simple $1d$ problems, the DMRG calculation can be viewed as an effective lower-bound on the performance of variational circuit ansatzes with $n_b\leq 5$ bond qubits. 
We note that, while the DMRG energy error is significantly lower than the variational qMPS results in this example, i) this relies crucially on the tractability of $1d$ DMRG calculations which does not extend to more complicated problems in $2d$, and ii) we will show that the GMPS+U method can be readily scaled to achieve comparable accuracy with far-fewer variational parameters (see Fig.~\ref{fig:ErrvsB}).%, and it is in those contexts that the quantum approaches may realize an advantage over the classical DMRG method. Further, we will show that scaling block size in the GMPS+U approach enables the variational methods to achieve comparable accuracy to this $\chi=2^5$ DMRG calculation with fewer variational parameters.

\begin{figure}[t] 
    \centering
    \includegraphics[width=0.5\textwidth]{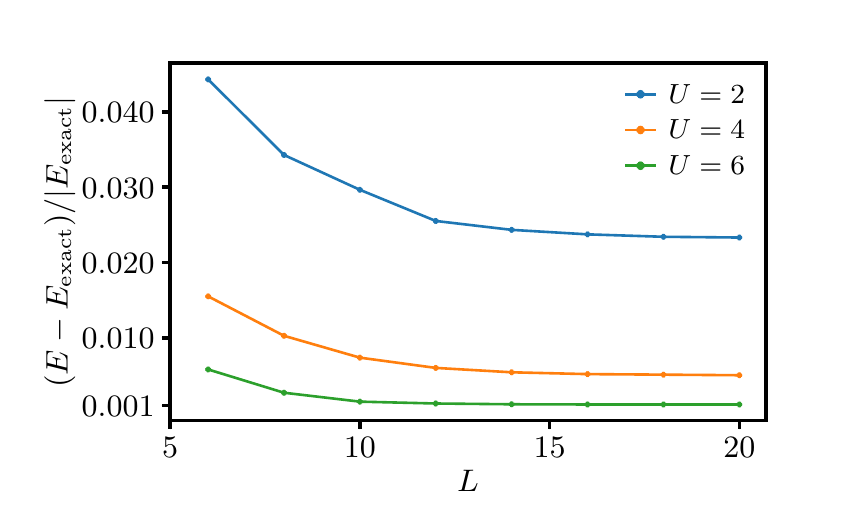}
    \caption{ {\bf Scaling of the GMPS+U Method when changing system size $L$ --}
    \label{fig:ErrvsL}
    the relative error in energy versus the total length $L$ of spinful sites for FH model at $U=2,4,6$ and $\nu = \frac{1}{2}$ and fixed block size $B=6$
        }
\end{figure}

To obtain a scalable variational ansatz, it is critical to reduce the number of variational parameters per site, $n_\text{var}$, required to achieve a desired accuracy.
Optimizing complex variational circuits with large $n_\text{var}$ is generically a classically difficult (non-linear, non-convex, and high-dimensional) problem
%\footnote{Evidence for this difficulty can be seen in the brick circuit data points with $\nu=\sfrac12$ and $U=5,6$, which have $\eps$ larger than that of the simpler GMPS+U ansatz despite having additional free parameters, suggesting trapping in a local minima of variational parameter space.}, 
and creates significant sampling overhead for measuring energies and gradients on quantum devices.
The complexity of the ansatzes ranges from $n_\text{var}=1$ for GMPS+J, $n_\text{var}=24$ for GMPS+U, and $n_\text{var}=80$ for the brick qMPS, to $n_\text{var} = \chi^2 = 1024$ for DMRG.
We note that, while we present simulation results for a finite number ($L=18$) of spinful sites, the algorithm complexity presented scales  (empirically) efficiently in $L$, and for the parameters explored this $L$ is sufficiently large to accurately capture the thermodynamic limit (see Fig.~\ref{fig:ErrvsL}).

In general, we observe that the GMPS+X techniques offer a significant reduction in the number of variational parameters $n_\text{var}$ required to achieve a given accuracy. At $\nu=\sfrac12$, the simplest ansatz, GMPS+J already achieves significant improvement over the mean-field results despite its extreme simplicity. Moreover, the GMPS+U ansatz outperforms the brick qMPS variational circuits across the entire $\nu, U$ parameter space explored, despite having significantly lower $n_\text{var}$.

The GMPS+U ansatz can be scaled to achieve higher accuracy by changing the block size $B$ (which also adjusts the number of variational parameters for building in correlations). Empirically, in the range of $B$ explored, we find that this allows one to adjust the complexity of the ansatz to achieve a desired target error rate. To explore the scalability, we focus on the weak Mott insulating regime ($U=2$, $\nu=\sfrac12$) where the GMPS+U error in Fig.~\ref{fig:vqe} is large, yet interactions are still important. Here, we observe a rapid decay of error with block size, that follows an approximately exponential trend $\eps\sim e^{-cB}$ with $c\approx0.7$ over the range of $B$ explored. 
We note that the mean-field GMPS compression error is already very low at $B=4$, and attribute the improvement with $B$  to the more flexible variational ansatz better capturing multi-particle correlations.
At the largest block sizes, $B=10$, the GMPS+U technique achieves performance equal to the classical $\chi=32$ DMRG despite having over an order of magnitude fewer variational parameters ($n_\text{var}=$54 versus 1024).
These results show that GMPS+U approach can achieve high-precision and scalable performance for simulating strongly-correlated electron models with far-fewer variational parameters than problem-agnostic circuit ansatzes.

%\blue{To address the possible concern about the finite size effect in the GMPS+U ansatz, we show how the relative error in energy scales with the system size. Here we show three examples $U=2,4,6$ at half-filling, as in the main text we show using $L=18$ spinful sites and block size $B=6$ it achieves reasonable accuracy, close to the optimized result from brick qMPS ansatz. Here we fix the block size and vary the system size $L$ and compare the relative error of the optimized energy at different $L$'s. From the result, the errors converge in system size $L$ and have already begun to saturate at $L=18$, the point we choose for the simulation.}

\subsection{Trapped-ion GMPS+X Demonstration}
We implement the simplest extended GMPS version, the GMPS+J variational ansatz, in the Quantinuum system model H1 trapped-ion quantum processor, focusing on the specific case of $\nu=\sfrac12$ and intermediate interaction strength ($U=4$), and for a Fermi-Hubbard chain of length $L=6$.
To avoid the lengthy process of hybrid classical-quantum optimization, we perform the optimization through classical simulation, and simply implement the classically-optimized circuit in hardware.
We employ a simple error-mitigation technique of post-selecting data on having the correct total number of particles (see Appendix \ref{app:em} for details).

\begin{figure}[t] 
    \centering
    \includegraphics[width=0.475\textwidth]{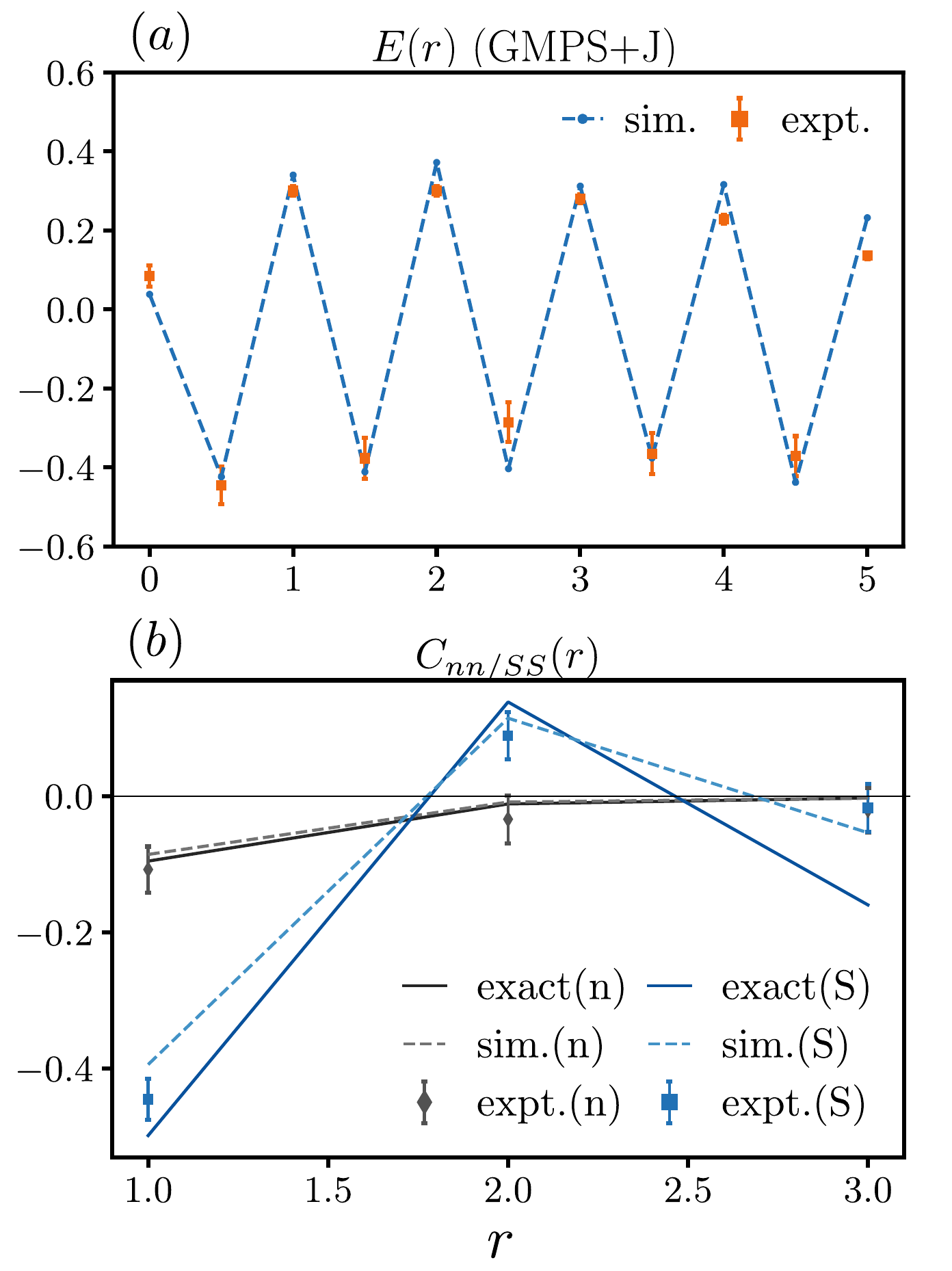}
    \caption{ {\bf  GMPS+J Hardware Implementation -- }
    \label{fig:FH_EX}
    for the Fermi-Hubbard chain at half-filling with U=4.
    (a) Energy density versus position in the GMPS+J ansatz (which is not explicitly translation invariant), with hopping energies shown at half-integer (bond-centered) positions.
    (b) Connected spin- (S) and charge-density- (n) correlators show spin-charge separation with rapid decay of charge and slower (antiferromagnetically modulated) decay of spin correlations.
        }
\end{figure}

 Fig.~\ref{fig:FH_EX} a,b respectively show the energy-densities (for each bond) and correlation functions for a chain of length $L=6$. Comparing to ideal (noiseless) circuit simulations, and exact (Bethe-ansatz) results, we find a good quantitive agreement to the experimental results within the statistical error-bars from a finite shot rate of between 400 and 1000 shots per data point (see Table~\ref{tab:shots} for details on the number of measurement shots, including error-mitigation post-selection).
 In addition to the quantitative agreement, the correlation data shows a clear separation of spin- and charge- with rapidly decaying charge-correlations and longer-range spin-correlations.

\section{Resource scaling of GMPS compression in 2d}
\label{app:2dgmps}
\begin{figure*}[t] 
    \centering
    \includegraphics[width=\textwidth]{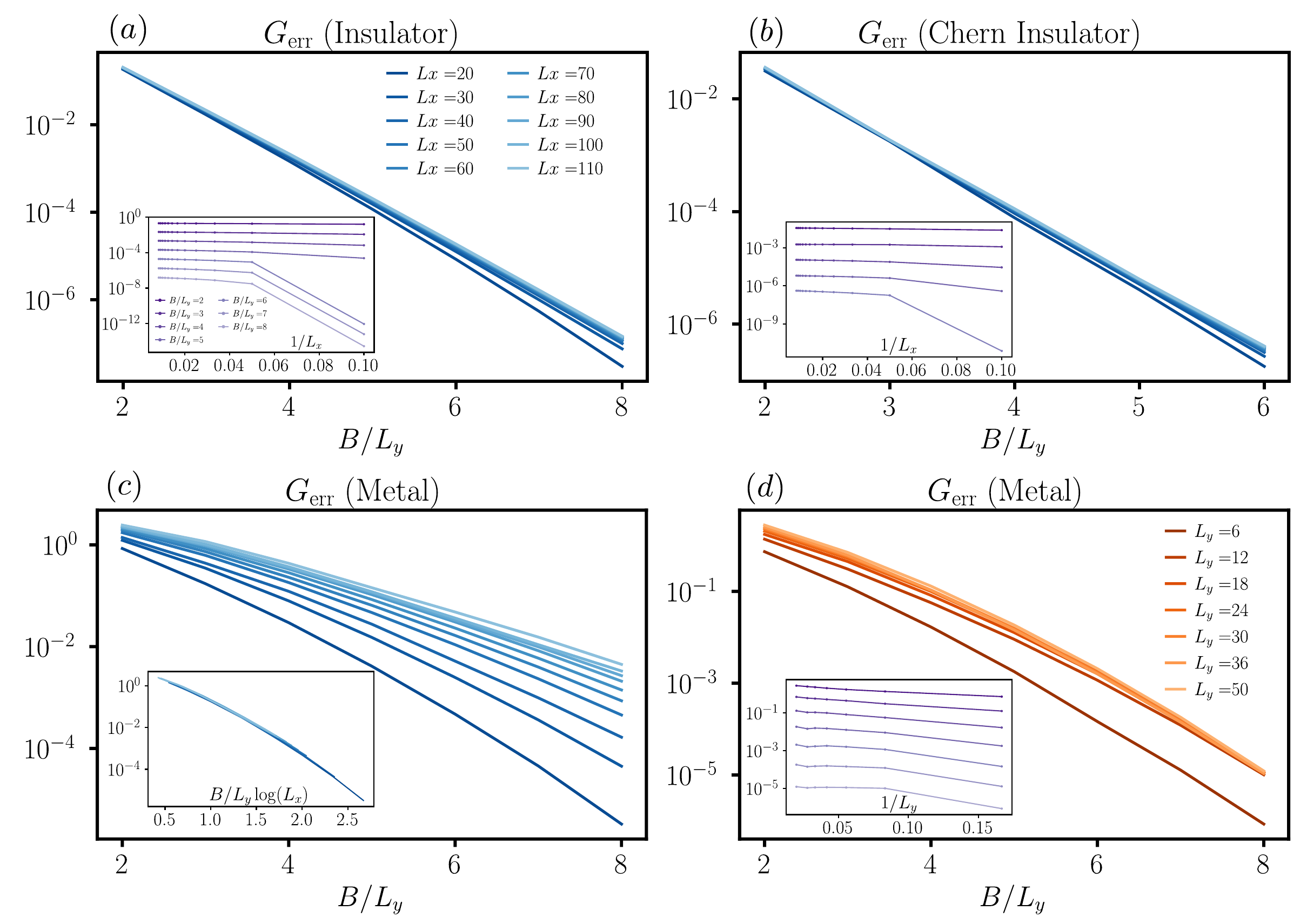}
    \caption{{\bf GMPS compression resource scaling}: (a)-(d) show the mean error $G_{\mathrm{err}}$ of Green's functions versus the GMPS compression block size for various systems and system sizes. $G_{\mathrm{err}}$ is defined by $G_{\mathrm{err}} = \sum_{i,j}|G^{c}_{i,j} - G^{o}_{i,j}|/V$, where $G^{o}$ and $G^{c}$ are the original and compressed Green's function and $V$ is the system size. In each plot we stop at the block size when the eigenvalues cannot be improved by increasing block size as it already reaches the limit of machine precision $10^{-15}$. (a)-(c) show the error versus the block sizes for different $L_x$ while $L_y$ is fixed (to 1,6,6); (d) shows the compression error versus the block sizes for different $L_y$ with $L_x=50$. The inset of (a), (b), (d) show the the $G_{\mathrm{err}}$ versus $1/L_x$ or $1/L_y$. The inset of (c) shows the $G_{\mathrm{err}}$ versus the block size re-scaled by $\log(L_x)$. 
    % In (a), the one-dimensional insulator ($L_y$=1) is constructed using SSH model with Hamiltonian $H_{SSH} = \sum^{\frac{N-1}{2}}_{i=1}(-t(1+\frac{\delta}{2}) a^{\dagger}_{2i-1}a_{2i} - t(1-\frac{\delta}{2}) a^{\dagger}_{2i} a_{2i+1} + h.c.)$ with $\delta=1$. In (b), the Chern insulator is constructed by the $\pi$-flux model in \cite{neupert2011fractional}, with parameter $t_1$=1, $t_2$=1 and the periodic boundary condition in $L_y$ direction. In (c) and (d), the two-dimensional metal is constructed by the tight-binding Hamiltonian $H=\sum_{\<i,j\>} (-t c^{\dagger}_i c_j + h.c.)$ where $\<\>$ denotes the neighboring sites, with the periodic boundary condition in $L_y$ direction.
    \label{fig:2dGMPS}
    }
\end{figure*}

While the GMPS method was originally developed for $1d$ systems, one can straightforwardly extend it to higher dimensions, by adopting an approach similar to that of $2d$-DMRG where a $2d$ system of length $L_x$ and width $L_y$ is considered as a stack of $L_x$ ``slices", or similarly by treating a $3d$ system as a stack of $2d$ cross-sections. 
Focusing on $2d$ (we expect similar results to hold in $3d$), we empirically investigate the resource scaling required to implement GMPS representations of three important classes of states~(see Fig.~\ref{fig:2dGMPS}):
\begin{enumerate}
    \item a topologically-trivial 1d band-insulator ($L_y$=1) is constructed from the Su-Schreifer-Heeger (SSH) model with Hamiltonian:
    $H_\text{SSH} = -t\sum_{j=1}^{N-1}\(1+(-1)^j\delta\)c^\dagger_jc^{\vphantom\dagger}_{j+1}+h.c.$ with $\delta=\sfrac12$,
    \item a topologically-nontrivial Chern-insulator is constructed from the ground-state of a square-lattice $\pi$-flux tight-binding model of \cite{neupert2011fractional}, with parameters $t_1$=1, $t_2$=1 and the periodic boundary condition in $L_y$ direction (to avoid gapless chiral edge states along the cylinder), and
    \item a two-dimensional metal with a Fermi-surface from an isotropic tight-binding is constructed from the ground-state of a square lattice tight-binding Hamiltonian $H=\sum_{\<i,j\>} (-t c^{\dagger}_i c_j + h.c.)$ where $\<\>$ denotes the neighboring sites, with the periodic boundary condition in $L_y$ direction
\end{enumerate}
%i) a trivial 2d band-insulator, ii) a topologically-nontrivial Chern-insulator, and iii) a metal with a Fermi surface (see Fig.~\ref{fig:2dGMPS}).
We measure the quality of the GMPS approximation via the mean error in the entries of $G_{ij}$ and focus on the regime where the eigenvalues obtained in the GMPS approximation can be improved until machine precision by using larger block sizes. In each case, we provide numerical evidence that the block size required for GMPS compression scheme follows that expected by the entanglement structure of these states, namely to achieve error $\eps$ requires $B\sim L_y\log \(1/\epsilon\)$ for both topological and trivial insulating states, and $B\sim L_y\log L_x \log\(1/\epsilon\)$ for metals, which is consistent with $B\sim S(L_x,L_y)$ where $S(L_x,L_y)$ is the half-system bipartite entanglement entropy. 

The Chern insulator result may initially seem surprising since the non-zero Chern number provides a fundamental obstruction to forming a localized Wannier basis, and at first glance, the GMPS algorithm may appear to be constructing such a basis. However, we note two points. First, the quasi-1d GMPS circuit structure only has locality along $x$, whereas the Chern obstruction only forbids simultaneous localization in both $x$ and $y$ directions. Specifically, the projection of the spatial coordinate into the orbitals of a Chern band  $(\hat{X},\hat{Y})$ fails to commute $\[X,Y\]\sim C$ where $C$ is the Chern number. Second, we have seen the GMPS circuit with $B\sim \log L_x$ is accurate even for metals with Fermi surfaces where the Wannier orbitals have algebraic decay in space. This indicates that the GMPS circuit \emph{cannot} simply be understood as a mapping to a strictly local Wannier basis. Technically, the unitary basis transformation implemented by the GMPS circuit is an upper triangular matrix and can produce long-range tails in the later entries.
%\red{Actually the transformation of basis is an upper triangular matrix, so there are very long tails with the later entries.} 

%, and since the GMPS algorithm essentially finds a strictly-local-approximation to a Wannier-basis of occupied states. However, we note that the quasi-1d structure we consider only ensures localization of along the $x$-direction, and may produce extended orbitals in $y$, whereas the Chern obstruction can be thought of as a Heisenberg-uncertainty principle for localizing simultaneously in both the $x$ and $y$ directions~\footnote{Specifically, the projection of the spatial coordinate into the orbitals of a Chern band  $(\hat{X},\hat{Y})$ fail to commute $\[X,Y\]\sim C$ where $C$ is the Chern number.}, and is compatible with this structure.

One significant consequence of this result is that it implies a polynomial advantage of qMPS techniques for simulating correlated Chern insulators compared to standard quantum simulation protocols (e.g. using adiabatic state preparation). 
Assuming that the numerically-established trends shown in Fig.~\ref{fig:2dGMPS} hold, then by leveraging i) standard rigorous results about adiabatic evolution~\cite{hastings2005quasiadiabatic} and trotterizing continuous time-evolution into circuit evolution~\cite{lloyd1996universal,childs2021theory}, and ii) previously analyzed~\cite{foss2021holographic} and experimentally feasible~\cite{chertkov2021holographic} methods for holographically implementing time evolution, we can establish an upper-bound on the qubit and gate resources required to prepare correlated Chern insulator (CI).
Specifically, the number of gates required to prepare a non-interacting CI ground-state via the GMPS method is $g\sim BL_x\sim L_yL_x \sim L^2$. Further, any correlated CI with integer Chern number is adiabatically connected to the non-interacting limit, and hence can be reached with constant-depth local time evolution, that can be implemented with constant qubit overhead via holographic time-evolution~\cite{foss2021holographic}. This GMPS+adiabatic-preparation approach has no free variational parameters and hence this performance bound does not rely on any assumption about the efficiency of optimizing variational circuits. By contrast, adiabatic state preparation of the Chern insulator from an un-entangled product state would inevitably require crossing through a phase transition since the product state and Chern insulator are topologically distinct phases (here we assume that the adiabatic evolution is performed with a local Hamiltonian). If this phase transition is continuous (second-order) with dynamical critical exponent $z$, then the minimal gap is $\Delta \sim 1/L^{z}$, which requires gate count $\sim L^2/\Delta^2\sim L^{2+2z}$. For physical transitions, $z\geq 1$, which places a bound $g\sim L^4$ for standard adiabatic preparation. Crossing a first-order transition would require much longer (exponential-in-L) adiabatic time.

Lastly, we remark that while we have focused on $2d$ systems here, going to higher dimensions does not add any qualitatively new ingredients, and we expect the trend to continue in $3d$, i.e. that the number of qubits in the GMPS continues to scale like the bipartite entanglement through a cross-sectional slice in all dimensions.
For example, we can trivially confirm this expectation for the special case of translationally invariant systems with periodic boundary conditions in $x,y$ and arbitrary boundary conditions in $z$, which can be reduced to a decoupled set of $\sim L_xL_y$ $1d$ systems along $z$ for each $k_x,k_y$, each of which can be compressed into a GMPS with a constant (gapped systems) or $\sim \log L_z$ scaling of qubits (metals).

%We also want to emphasize that although the numerical simulations are done in 2d systems, similar arguments about the GMPS compression error scaling can be made in even higher systems like 3d systems and so on. This is because the encoding is the same for higher-dimension systems. 2d systems are encoded into GMPS by being treated as a stack of 1d slices, which can be applied to 3d systems and in general higher dimension systems as well. The increase of dimension is then reflected in the $L^{d-1}$ term in the scaling we have stated.

\section{Discussion}
In this work, we introduced and experimentally demonstrated 
%two complementary 
holographic approaches to prepare ground-states of correlated electron systems.
%relevant that 
%We leveraged the Gaussian state compression scheme of Fishman and White, qMPS encodings, mid-circuit measurement, and qubit reuse to simulate models with many orbitals with few hardware qubits and modest gate resources.
%The accuracy of the results we obtain using at most 6 data qubits, would require dozens of qubits using conventional (non-holographic) techniques. 
The efficient preparation of gapped and gapless mean-field ground states as qMPS (obtained by our qMPS adaptation of the GMPS compression scheme of Fishman and White), in conjunction with holographic time-evolution algorithms~\cite{foss2021holographic}, formally establishes that qMPS with efficient circuit resources (i.e. scaling polynomially with system size) can accurately capture any state that is continuously connected to a mean-field fermion state (possibly via a continuous phase transition). 
This includes most phases of matter relevant to practical material simulations, such as metals, correlated trivial- and topological- insulators, magnets, superconductors, density-wave states, et cetera. 

Compared to standard adiabatic state preparation protocols, the combination of the holographic approach, qMPS encoding and Gaussian compression bring important advantages that are already apparent in the mean-field state preparation.
For example, 
for gapped Hamiltonians, the holographic qMPS encoding offers a polynomial reduction in qubit resources (from $q\sim L^d$ to $q\sim L^{d-1}$), and for $2d$ topological systems with non-trivial Chern number further allows a polynomial reduction in gate count (from $g\sim L^4$ to $g\sim L^2$)~\footnote{These gate count scalings are for \emph{local} quasi-adiabatic preparation~\cite{hastings2005quasiadiabatic} sufficient to accurately reproduce correlations of local operators, true adiabatic preparation to produce full many-body state fidelity close to $1$ requires a multiplicative factor of $L^2$.}. Similarly, 
compared to the Givens rotation approach
%technique 
used in recent hardware demonstrations of Hartree-Fock ground-state preparation~\cite{arute2020hartree} 
%on comparable-length (but spinless) Fermion chains
which required $q\sim L$ qubits and $g\sim L^2$ gates, qMPS constructed via GMPS compression require only $q\sim\log L \log\eps^{-1}$ and $g\sim L\log L\log \eps^{-1}$, achieving a polynomial reduction in both resources~\footnote{We note that the gate count gains are asymptotic, for the task of simulating a spinful $L_x=8$ FH chain, both methods require nearly the same number of entangling gates ($130$ M\o lmer-S\o rensen gates for GMPS versus $128$ $\sqrt{\rm iSWAP}$ gates for given rotations). For any $L_x>8$, the GMPS approach requires fewer gates.}. Previously, the most efficient Gaussian state preparation protocol
was the fermionic Fast Fourier transform (FFFT), which similarly uses $g\sim L\log L$~\cite{ferris2014fourier} gates; however, unlike the GMPS technique, the FFFT is restricted to 
%fermionic fast fourier transform (FFFT) methods can achieve similar asymptotic performance of $g\sim L\log L$~\cite{ferris2014fourier}, however this requires long-range gates that spoil the holographic savings in qubit number (and may be expensive to implement in many architectures). Additionally, the FFFT method is specific to preparation of 
translation-invariant (plane-wave-like) states and requires long-range gates that spoil the holographic savings in qubit number (and may be costly to implement in many architectures). The source of the efficiency of Gaussian compression is the exploitation of the near-area-law entanglement behavior of physical gapped and gapless ground-states. Retaining this structure in the qMPS then allows for compact preparation of near-area-law interacting ground-states. We demonstrated this here by preparing ground-states of interacting 1D fermion Hamiltonians, including those with extended power-law ground-state correlations in $G_{ij}$, using  GMPS+X circuits adapted from the mean-field circuits.

Natural targets for follow-on work include tackling higher-dimensional systems via 2$d$ or 3$d$ qMPS or qTNS techniques, exploring alternative fermion-to-qubit encodings that are well-suited to qTNS methods, and exploring qTNS-based embedding techniques~\cite{haghshenas2021variational} to simulate realistic material models with complex (e.g. long-range) interactions. 

Given the importance of noise and errors in near-term implementations, it will further be important to systematically assess the impact of gate errors on GMPS-based circuit ansatzes. For example, related work on holographic multi-scale entanglement renormalization ansatzes (MERA) suggests that holographic algorithms possess an intrinsic degree of noise-resilience~\cite{kim2017robust}. For qMPS circuits describing gapped $1d$ systems, one expects a spectral gap in the MPS transfer matrix, such that perturbing the circuits weakly with gate errors, would lead to a small, finite correction to the qMPS steady-state properties even in the limit of infinitely long systems ($L\rightarrow \infty$). Exploring the systematic dependence of this noise susceptibility for different physical types of states is a potentially interesting target for future work, and the analytic tractability of free-fermion systems may provide a tractable set of examples to explore the systematic dependence of noise sensitivity of qMPS based on the entanglement structure of the ground-state being approximated.

\vspace{4pt}\noindent{\it Acknowledgements -- } We thank Itamar Kimchi, Roger Mong, and Michael Zaletel for insightful conversations.
We acknowledge support from NSF award DMR-2038032 (YZ, AP), NSF-Converence Accelerator Track C award DMR- (DN, GKC), from the Alfred P. Sloan Foundation through a Sloan Research Fellowship (AP). RH was supported by the US Department of Energy, Office of Science, via Award No. DE-SC0019374. Additional support from GKC was provided by the Simons Collaboration on the Many-electron Problem and the Simons Investigatorship. This research was undertaken thanks, in part, to funding from the Max Planck-UBC-UTokyo Center for Quantum Materials and the Canada First Research Excellence Fund, Quantum Materials and Future Technologies Program. Numerical calculations were performed using supercomputing resources at the Texas Advanced Computing Center (TACC). 

\appendix
%\onecolumngrid
%\pagebreak
%\twocolumngrid

\section{Quantinuum QCCD Architecture}
\label{appendix:circuit}
Hardware implementations are performed on Quantinuum's System Model H1 trapped-ion quantum processor~\cite{pino2021demonstration}, which uses a quantum charge-coupled device (QCCD) architecture, based on a Quantinuum-fabricated planar chip trap operating with three parallel gate zones and 10 qubit ions. One-qubit ($1q$) gates implement $\pi/2$ rotations about an arbitrary axis in the $\sigma^{xy}$-plane. The native entangling two-qubit gate is a M\o lmer-S\o rensen gate wrapped with single-qubit dressing pulses to achieve a phase-insensitive operation $u_{MS}=\exp[-i\frac{\pi}{4}Z\otimes Z]$ \cite{pino2021demonstration}. The one-qubit gates and two-qubit gates have typical average infidelities (determined by randomized benchmarking techniques) of: $\epsilon_{1q}\approx10^{-4}$ and $\epsilon_{2q}\approx2-5\times 10^{-3}$.

Due to the large sampling overhead and relatively low clock speed of the QCCD device, we chose to only perform part of the holographic variational quantum eigensolver (holoVQE) algorithms on the quantum device.
Namely, in all cases, we performed the variational optimization of circuit parameters through classical simulations and implemented only the optimized circuit in hardware. 
While this procedure does not address the effect of hardware noise on the variational optimization (nor allow for possible variational cancellation of coherent errors), it nevertheless allows one to test how realistic hardware errors affect the achievable variational errors with different circuit types.

\section{Details of GMPS Circuit implementations}
\label{appendix:design}
In the GMPS compression algorithm, a basis transformation specified from the eigenvector of the Green's function block transfers the least entangled states to the first site of the block. The corresponding basis transformation is decomposed into a series of nearest neighbor two-site gates which rotate the Green's function block into the occupation basis. The circuit and gate parameters are obtained from the eigenvectors.

Rotating the first site to the eigenvector $(v_1, v_2, ..., v_{B-1}, v_{B})$ requires $(B-1)$ two-site gates in total.
%to complete the basis transformation. 
The first gate acts on site $(B-1, B)$ and is labeled as $V_{B-1}$. $V_{B-1}$ satisfies $v^{T} V_{B-1} = (v_1, ... v'_{B-1}, 0)$. In general the gate $V_i$ satisfies $(v_1,...v'_{i+1},0,...0)^{T} V_i = (v_1,...,v'_{i},0,...0)$ and it takes the form:
\begin{equation}
V_i = V(\theta_i) = 
\begin{pmatrix}
\cos\theta_i & -\sin\theta_i \exp{(i\phi_i)}\\
\sin\theta_i\exp{(-i\phi_i)} & \cos\theta_i\\
\end{pmatrix}
\label{eq:v}
\end{equation}
When the Green's function is real, $\phi = 0$ and $\theta_{B-1} = \tan^{-1}(v_B/v_{B-1})$, thus the gate $V_i$ takes the same form as in Ref. ~\cite{fishman2015compression} When the Green's function is complex in certain cases (like in Chern insulators), the eigenvectors are also complex thus the extra phase factor $\phi$ becomes necessary.
The entries of the Green's function are updated once the gate acts on it. The next gate acts on $(B-2, B-1)$ and so on. This procedure gives us $V_{B-1}$, $V_{B-2}$,...,$V_1$. The total unitary transformation is $V_{\mathcal{B}_1} = V_{B-1} V_{B-2} ... V_1$. Acting this on the Green's function $G$ yields the transformed Green's function $V^{\dagger}_{\mathcal{B}_1} G V_{\mathcal{B}_1}$ with $n_1\approx 0 $ or $1$.
The procedure is repeated for sites 2,... $B+1$ to obtain $V_{\mathcal{B}_2}$. For the last few sites, the block size becomes the number of the remaining sites. In the end we obtain the total unitary transformation $V=V_{\mathcal{B}_1} V_{\mathcal{B}_2} ... V_{\mathcal{B}_{N-1}}$. 

\section{Variational Circuit Architectures}
\label{app:vqe}
This subsection details the architecture for the variational circuits shown in Fig.~\ref{fig:vqe}. This includes two types of GMPS+X circuits that augment the GMPS Hartree-Fock circuit with additional variational circuits: \\
i) the GMPS+J ansatz (where `H' stands for `Heisenberg') depicted in Fig.~\ref{fig:vqe}a) geared towards large-$U$ and half-filling, and \\
ii) the GMPS+U ansatz (Fig.~\ref{fig:vqe}b) which is agnostic to filling and $U$ (although it tends to perform best at large or small $U$), and \\
iii) a problem-agnostic brick circuit (Fig.~\ref{fig:vqe}c), which can achieve high accuracy, but at the cost of introducing a comparatively large number of variational parameters.

\subsection{GMPS+J}
The GMPS+J ansatz exploits the physical picture that the half-filled Fermi-Hubbard model behaves like a Heisenberg spin-chain at large-$U$'s, by introducing a variational ``pre-entangling" circuit before the GMPS circuitry (Fig.~\ref{fig:vqe}a) using a single extra bond-qubit and layer of variational gates to build in quantum fluctuations of the AFM spin-texture in the Hartree-Fock ground-state.
The aim is to emulate a qMPS state of the electron spins without altering the charge state.
In our approach, we encode each spinful fermion site into two qubits whose computational basis encodes the particle occupation number of the up and down spin orbitals: $|n_\up,n_\down\>$.
The input state to the GMPS circuit is simply a product state of Fermi-Hubbard sites with a single particle on each site, and with spin alternating between up and down.
To entangle the spin-degrees of freedom on different sites without affecting their charge, we employ a basis transformation using a CNOT gate, which corresponds to the following map:
\begin{align}
|n_\up, n_\down\> &\longrightarrow |\text{flavor}, 2P_F-1\>:\nonumber\\
|0, 0\> &\longrightarrow |0,0\> \nonumber\\
|0, 1\> &\longrightarrow |0,1\> \nonumber\\
|1, 0\> &\longrightarrow |1,1\> \nonumber\\
|1, 1\> &\longrightarrow |1,0\>.
\end{align}
In the transformed basis the second qubit encodes the even/odd fermion parity $P_F = (-1)^{n_\up+n_\down}$. The flavor qubit's interpretation depends on the value of $P_F$. When $P_F$ is odd, there is a single electron per site and the first qubit's Pauli operators correspond to the electron spin operator, e.g. $Z_1=c^\dagger_s\sigma^z_{ss'} c^{\vphantom\dagger}_{s'}$ (and similarly for $Y,Z$). While our initial states always have odd fermion parity, for completeness, we mention that for even fermion parity the ``flavor" qubit encodes the occupation number of a spin-singlet Cooper pair created by $c^\dagger_\up c^\dagger_\down$.
%$(n_\uparrow,\n_\downarrow)$ 
%

In the flavor/parity basis, the electron-spin on each site now maps to the state of a single qubit, and we can apply a variational $XX+YY$ gate
\begin{align}
    u_{\rm XX+YY}(\theta)=\exp\[-i\frac{\theta}{2}(X\otimes X + Y\otimes Y)\]    
\end{align}
between this flavor qubit and an extra bond qubit, emulating the variational circuitry for a qMPS approximation to the Heisenberg spin-chain ground-state explored in~\cite{foss2021holographic}.
We then transform back to the original particle occupancy basis to implement the GMPS circuit for the Hartree-Fock ground-state. 
The GMPS+J circuit introduces only a single extra variational parameter per site.
%
% \begin{figure}[t] 
%     \centering
%     \includegraphics[width=0.2\textwidth]{figures/FigA1_CorrCircuit.pdf}
%     \caption{ {\bf Circuit to build in correlations}: the schematic circuit of basis transformation and $XX+YY$ gates for adding in correlations before the free fermion compression circuit
%     \label{fig:CorrCircuit} 
%     }
% \end{figure}

%\subsection{GMPS+U}
%In the GMPS+U ansatz, we first add an additional variable-rotation $u_{zz}(\theta_i) = e^{-\frac{i}{2}\theta_i Z\otimes Z}$ gate to each of the free-fermion GMPS gates, with independent variational parameters $\theta_i$ for each gate. We then add generic one-qubit gates $R$ with the form:
%\begin{equation}
%R(\alpha, \beta,\gamma) = 
%\begin{pmatrix}
%\cos(\frac{\alpha}{2}) & -e^{i\gamma}\sin(\frac{\alpha}{2}) \\
%-e^{i\beta}\sin(\frac{\alpha}{2}) & e^{i(\beta+\gamma)}\cos(\frac{\alpha}{2})\\
%\end{pmatrix}
%\end{equation}
%to exploit the full freedom in the ansatz. These gates are non-Gaussian for generic values of $\alpha$, and produce a correlated non-Gaussian fermion qMPS. The number of variational parameters in this ansatz is four times the number of two-qubit gates  ($\sim N_{\rm o}B$) in the GMPS circuit.

\subsection{Brick circuit qMPS}
Besides the GMPS+X approaches, we also employ a problem-agnostic variational circuit approach, which we will label by the term brick qMPS. Specifically, we fix a brickwork circuit architecture with general particle-number conserving two-qubit gates (see Fig.~\ref{fig:vqe}) to generate the tensors of a qMPS with $\chi=2^4$ and classically minimize $\<H_\text{FH}\>$ with respect to the variational parameters using quimb~\cite{gray2018quimb}. 
The number of bond-qubits and gates in this ansatz can be arbitrarily scaled to achieve larger expressiveness. However, this introduces a large number of variational parameters ($5$ parameters per two-qubit gate), which may be difficult to train on larger problem instances (e.g. $2d$ models).

Our brickwork circuit ansatz employs general particle number-conserving two-qubit gates to enable noise mitigation based on post-selecting the data on having the correct total particle number, with unitary:
\begin{align}
&U_{2q}(\gamma,\phi,\zeta,\chi,\theta) = 
\nonumber\\
&\begin{pmatrix}
e^{i(\gamma+\phi)} & 0 & 0 & 0 \\
0 & e^{i(-\gamma+\phi+\zeta)}\sin\theta & e^{-i(\chi+\gamma+\phi)}\cos\theta & 0\\
0 & e^{i(\chi-\gamma+\phi)}\cos\theta & e^{-i(\gamma+\phi+\zeta)}\sin\theta & 0\\
0 & 0 & 0 & e^{i(\gamma - \phi)}\\
\end{pmatrix}
\end{align}
where $(\gamma, \phi, \zeta, \chi, \theta)$ are variational parameters (independently chosen for each two-qubit gate in Fig.~\ref{fig:vqe}c).

\section{Error mitigation}
\label{app:em}
\begin{figure}[t] 
    \centering
    \includegraphics[width=0.4\textwidth]{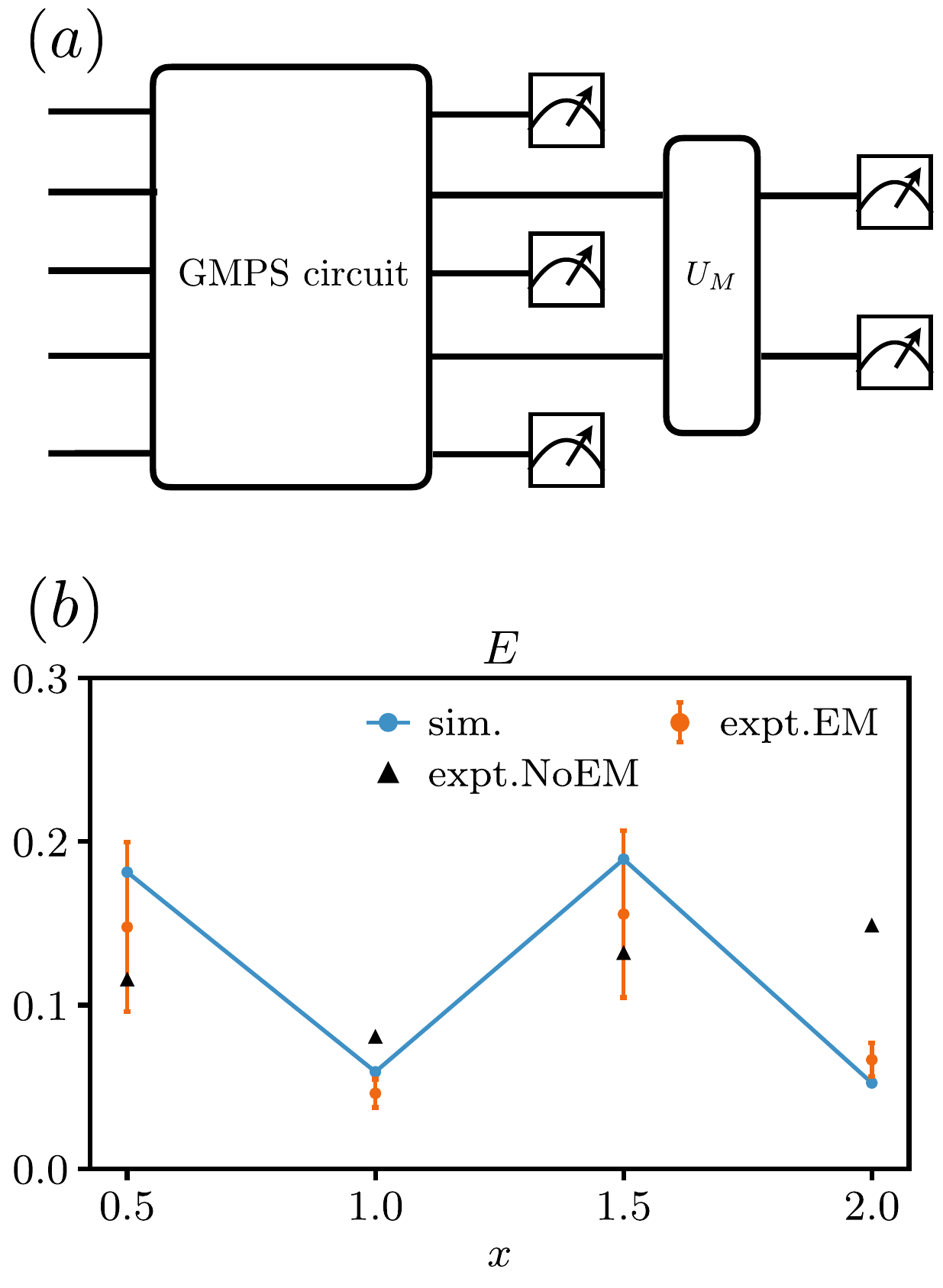}
    \caption{{\bf Error mitigation--} (a) To mitigate hardware errors, we post-select on results with the correct total particle number. To simultaneously measure total particle number and Green's function elements $\<c^\dagger_i c_j\>$, we add a basis transformation gate $U_M$ for the corresponding qubits of the Green's functions before measuring them and each qubit is measured in the Pauli Z basis as in Ref.~\cite{arute2020hartree}. In the holographic implementation, the first qubit being measured cannot be reset and reused until both sites are $i,j$ are measured. (b) A comparison between measurement results for qMPS VQE method of U=6, half-filled Fermi-Hubbard model with (EM) and without (NoEM) the error mitigation. Details of post-selection success rate and shot counts are shown in Table~\ref{tab:shots}
    \label{fig:NoiseMiti}}
\end{figure}

\begin{table*}[]
\def\arraystretch{1.3}
\begin{tabular}{llc*{6}{c}r}
terms: && $c^{\dagger}_{\uparrow 0} c_{\uparrow 1}$ & $c^{\dagger}_{\downarrow 0} c_{\downarrow 1}$ & $c^{\dagger}_{\uparrow 1} c_{\uparrow 2}$ & $c^{\dagger}_{\downarrow 1} c_{\downarrow 2}$ & $c^{\dagger}_{\uparrow 2} c_{\uparrow 3}$& $c^{\dagger}_{\downarrow 2} c_{\downarrow 3}$& $n_{i\uparrow}n_{i\downarrow}$ & \\
\hline
U & $\nu$~~~~~              & total shots & & & & & &  & success rate\\
\hline
$6$ & $\sfrac12$ & 400 & 400 & 400 & 400 &  &  &  1000 & 58.5\% \\
$4$ & $\sfrac12$ & 1000 & 1000 &  &  &  &  & 1000 & 66.4\% \\
$1$ & $\sfrac12$ & 400 & 400 & 400 & 400 &  &  & 1000 & 42.1\% \\
$6$ & $\sfrac13$ & 800 & 800 & 800 & 800 &  600 & 800 &  1000 & 42.1\% \\
$4$ & $\sfrac13$ & 800 & 800 & 1000 & 1000 & 1000 & 1000 &  1000 & 39.1\% \\
$1$ & $\sfrac13$ & 800 & 1000 & 800 & 1200 &  1000 & 1000 &  1000 & 29.8\%\\
\end{tabular}
\caption{Total shots for different measurements ($i = 1,2,3$ for $\sfrac{1}{3}$-filling states and $i=1,2$ for $\sfrac{1}{2}$-filling states), success rate defines the proportional of total shots kept after the noise mitigation post selection}
\label{tab:shots}
\end{table*}
%
%\subsubsection{Particle-number post-selection}
As a simple noise mitigation method, we post-select our data on having the correct total particle number. Since a noiseless implementation of the circuits conserves total particle number, any deviation from the ideal number can only be caused by gate errors (though not vice versa). This symmetry-based postselection gives a modest but noticeable improvement in measuring Green's functions and density correlations. An example is shown in Fig.~\ref{fig:NoiseMiti}, comparing the measurement results with and without the noise mitigation.

When all the measurements are in the Pauli-Z bases, it is convenient to keep track of the particle number just using the measurement results. The most straightforward way of measuring Green's functions $\<c^{\dagger}_{r_1} c_{r_2}\>$, would be to implement separate measurements of its real and imaginary parts as Pauli strings: "...XZ...Z...ZX..." and "...YZ...Z...ZY...". However, as these strings do not individually commute with total particle number, the method would be incompatible with our error mitigation scheme. Instead, we add additional gates to map $X_i X_j + Y_iY_j$ and $Z_i+Z_j$ eigenstates to the computational basis so that the real part of $G$ and the total number can be simultaneously measured. This mapping is achieved by the unitary (written in the $Z_i,Z_j$ eigenbasis):
\begin{align}
U_M =
\begin{pmatrix}
1 & 0& 0& 0 \\
 0& \frac{1}{\sqrt{2}}& \frac{1}{\sqrt{2}}&0 \\
 0& \frac{1}{\sqrt{2}}& -\frac{1}{\sqrt{2}}&0 \\
 0& 0& 0& 1
\end{pmatrix}.
\end{align}
%^{\vphantom\dagger}

Since the state is built sequentially from left to right, performing this rotation requires postponing the mid-circuit measurement and reset of the qubit at $r_1$ until site $r_2$ is reached, which requires a single extra hardware qubit compared to the basic qMPS circuit without error mitigation.

\section{GMPS for superconductors and thermal states}
\label{appendix:non-conserving}
While Ref.~\cite{fishman2015compression} focused on the classical GMPS method for ground-states of real, number conserving free-fermion Hamiltonians, it is straightforward to generalize this framework to general Gaussian fermion pure- and mixed- states. The case of complex Hamiltonian entries is already accounted for by the phases, $\phi$ in equation Eq.~\ref{eq:v} above. Below we briefly sketch the generalizations for superconducting and mixed (e.g. thermal) Gaussian states.

\subsection{Particle Non-Conserving States (Superconductors)}
For superconducting mean-field states that do not conserve particle number, a standard convenient trick is to redundantly represent $N$ fermion orbitals using a $2N$-component Nambu operator:
\begin{align}
\psi_{i,s} = \begin{pmatrix} c^{\vphantom\dagger}_1, & \dots c^{\vphantom\dagger}_N, & c_1^\dagger, & \dots & c_N^\dagger \end{pmatrix}^T
\end{align}
where $s = +1$ corresponds to the particle ($c_i$) block and $s=-1$ to the hole ($c_i^\dagger$) block. Denoting the Pauli matrices in this particle-hole space as $\vec{\tau}_{ss'}$, this description has a particle-hole redundancy $\psi^\dagger = \tau^1\psi$. The classical part of the GMPS algorithm can then be run as normal in this Nambu basis, except that each time one identifies an approximate block eigenvector $\v{v}$ with block occupation number $n$ close to $0$ or $1$, its particle hole conjugate $\tau^1\vec{v}$ will also be a block eigenvector with the block occupation number $1-n$, which is equally well localized to the block. These vectors should be simultaneously decoupled from the rest of the system by implementing rotations $V_B$ and $\tau^1 V_B\tau^1$ on $G$. In the many-body language the many-body operator corresponding to single-particle rotation $V_{ij}$ is $e^{\psi^\dagger_{is} (\log V)_{is,js'} \psi_{js}}$, which will automatically perform both $V_B$ and $\tau^1 V_B\tau^1$ rotations due to the particle-hole redundancy.

\subsection{Thermal states via purifications}
It is also possible to prepare a purified version of Gaussian mixed states via GMPS methods at the cost of doubling the number of qubits required compared to a pure Gaussian state. Without loss of generality, we consider this method for thermal states of the form: $\rho = \frac{1}{Z}e^{-c^\dagger_ih_{ij}c_j}$ (where we have chosen normalization of $h$ such that temperature is $1$, and $Z$ normalizes $\text{tr}\rho=1$), since any Gaussian mixed state can be represented in this way. We expect the compression to be effective when $h$ is a local Hamiltonian, since these thermal states will have an area-law scaling of mutual information~\cite{wolf2008area} and efficient matrix-product density operator form~\cite{jarkovsky2020efficient}.

The basic idea is to prepare a thermofield-double (TFD) type state on a doubled system with the fermion creation operators of the system and double respectively labeled as $c_i$, $a_i$ (here `$a$' stands for ancilla). To start, consider just a single mode thermal state $\rho_T = \frac{1}{Z}e^{-\e c^\dagger c}$. This can be prepared as a TFD state: $|\Psi_\text{TFD}\>=\frac{1}{\sqrt{Z}}e^{-\e c^\dagger a/2}|0\>_s\otimes |1\>_a$, which has the properties i)  $\rho_T = \text{tr}_a |\Psi_\text{TFD}\>\<\Psi_\text{TFD}|$, and ii) $|\Psi_\text{TFD}\>$ is a Gaussian fermion state that can be approximately prepared as a GMPS acting on the doubled $\{c,a\}$ system. For multiple modes, this simply generalizes to $|\Psi_\text{TFD}\> = \frac{1}{\sqrt{Z}}e^{-c^\dagger_i h_{ij} a_j}|0\>_s \otimes |1\>_a$ where $|0\>$ is the all-empty state, and $|1\>$ is the all-full state respectively (as can be seen by working in the eigenbasis of $h_{ij}$ which reduces to the single-mode problem above).

This Gaussian TFD (GTFD) preparation could be particularly effective as a starting point for variational thermal state preparation schemes based on minimizing the free energy $F=\<H\>-TS$ where $S=-\tr\rho_s\log\rho_s$. Namely, whereas computation of $\<H\>$ for a variational state is straightforward on a quantum computer, measurements of $S$ for an unknown state incur exponential sampling overhead~\cite{islam2015measuring,brydges2019probing}. However, if one starts with a GTFD state, and adds subsequent variational circuit layers acting on the $c$ system alone, the entropy of the $c$-system remains that of the initial thermal state, which can be efficiently calculated classically. The resulting state has a fine-tuned entanglement spectrum that is the direct product of many independent two-state systems (one per fermion orbital). Generic thermal states instead exhibit random-matrix type entanglement spectrum with level-repulsion between nearby entanglement energies. However, recent work~\cite{martyn2019product} provides evidence and arguments that the fine details of the entanglement level spacing statistics are not visible in physical quantities of interest such as correlations of local observables and that product-state entanglement spectrum ansatzes are effective at reproducing such observables in correlated thermal states.

\begin{figure}[t]
\centering
    \includegraphics[width=0.5\textwidth]{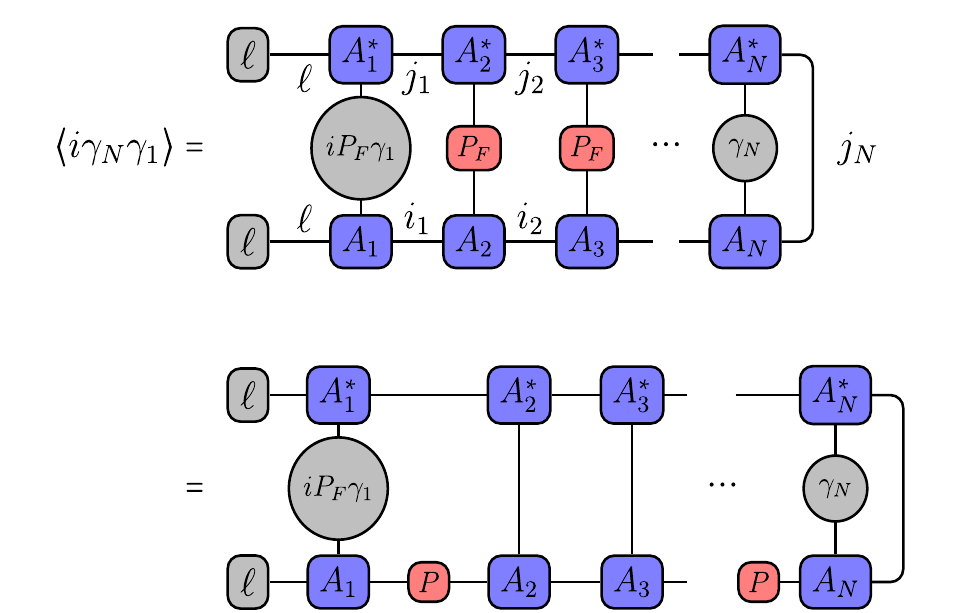}
    \caption{ {\bf } The schematic representation of applying equation (B1) to fermionic operator $\gamma_{1,N}$
    \label{fig:fh2} 
    }
\end{figure}
\begin{figure*}[] 
    \centering
    \includegraphics[width=\textwidth]{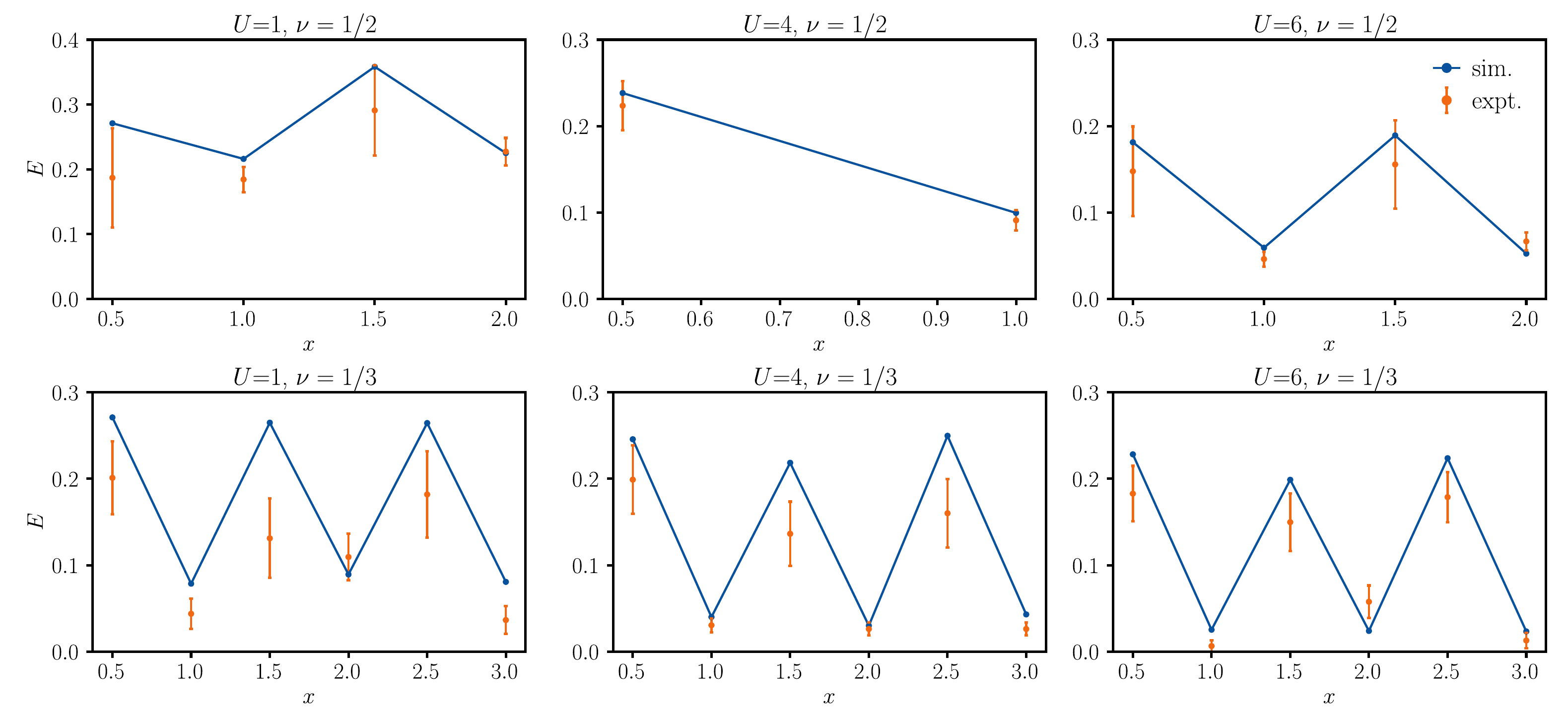}
    \caption{{\bf brick qMPS method measurement data} of Green's functions and density correlations of the Fermi-Hubbard model of different U's and fillings. Absolute value of hopping energies $t\<\frac{1}{2}(c^{\dagger}_{i\uparrow} c_{i+1 \uparrow} + c^{\dagger}_{i\downarrow} c_{i+1 \downarrow})\>$ are shown at bond-centered  coordinates with half-integer positions $x=i-0.5$, where $i$ labels the spinful sites. On-site repulsion density correlations $\<n_{i \uparrow} n_{i \downarrow}\>$ are shown at site-centered coordinates (integer $i$).
    \label{fig:fh} 
    }
\end{figure*}

\section{Circumventing Jordan-Wigner String Measurements}
\label{appendix:noJW}
Measuring fermion correlations between distant sites requires measuring long Jordan-Wigner (JW) strings. Errors in measurement will generically cause such observables to decay exponentially in distance. While this effect is minor for moderate system sizes explored in this paper, it may become a dominant source of error in larger-scale models. Here, we show that in holographic qMPS simulations, it is possible to avoid the measurement of long JW strings, by ``pulling" the strings back into the virtual bond-space of the MPS (see Appendix A of \cite{corboz2009fermionic}) where they largely cancel except for additional boundary terms that can be measured by sampling results from a small number of different qMPS contractions.

The measurement of a Jordan-Wigner string can be simplified in the following way: first, we pull back the Jordan-Wigner strings onto the bond space. Specifically, for tensors that have definite fermion parity, it is always possible to pull back the action of fermion parity operators: $P_{F,i} = (-1)^{n_{F,i}}$ where $n_{F,i}$ is the total number of physical fermions on site $i$, which appears in the JW string, into operators acting on the input and output bonds of the tensor:
\begin{equation}
(P_F)_s A_s^{ij} = P^{\dagger}_{ik} A_s^{ij} P_{lj}
\end{equation}
where $P$ is the representation of fermion parity on the bond space. 
%Restricting our attention to non-projective representations of fermion parity on the bond-space (i.e. neglecting the case of a topological superconductor with Majorana edge states) we can choose $P$ as a diagonal operator: $P = \sum _{i=1}^{\chi}(-1)^{|i|}|i\>\<i|$. 
Crucially, each bond in the middle of the JW string has a $P$ from the tensor to its left and a $P^\dagger$ from the tensor to its right, which cancel, leaving only $P$'s at the terminal bonds, as shown schematically in Fig.~\ref{fig:fh2}.

The transfer matrix with $P$ acting on the lower-leg but not the upper-leg is not a valid quantum channel, but we can decompose it as a linear combination of a small number of quantum channels that can be separately measured, and then linearly combined to compute the desired result. 
To this end, we introduce the following two basic operations on the bond space. First, denote measurement of bond-fermion parity $M_P$:
\begin{align}
M_P &= \Pi_{P=1}\otimes \Pi_{P=1} - \Pi_{P=-1}\otimes \Pi_{P=-1}  
\nonumber\\
	&= \frac{(1+P)}{2} \otimes \frac{(1+P)}{2} - \frac{(1-P)}{2} \otimes \frac{(1-P)}{2}
\nonumber\\
	&= \frac12\[P\otimes 1 + 1\otimes P\]
\end{align}
Implementing $M_P$ requires measuring $P$ on the bond register without collapsing the full bond-wave function, which in practice can be done using an ancilla and standard phase-kickback scheme (one of the physical qubits which has already been measured and is currently not active can play this role so that the total qubit resource requirements are unaffected).

Second, we define the operator $\sin(\frac{\pi}{4}\text{ad}_P)$ where $\text{ad}_P\circ = [P, \circ]$: 
\begin{align}
\sin\(\frac{\pi}{4}\text{ad}_P\) &= \frac{1}{2i}\(e^{i\frac{\pi}{4}\text{ad}_P} - e^{-i\frac{\pi}{4}\text{ad}_P}\)
\nonumber\\
	&= \frac12\(P\otimes 1 - 1\otimes P\),
\end{align}
and each of the terms: $e^{i\frac{\pi}{4}\text{ad}_P} = e^{\pm i\pi P/4}\otimes e^{\mp i\pi P/4}$ can be implemented simply by applying the unitary operator $e^{\pm i\pi P/4}$ to the bond-qubit register.

From these two ingredients, the desired operation of applying $P$ to the lower bond-legs but not the upper ones can then be written as:
\begin{align}
P\otimes 1 = M_P + \frac{1}{2i}\(e^{ i\pi P/4}\otimes e^{-i\pi P/4}-e^{- i\pi P/4}\otimes e^{ i\pi P/4}\)
%\frac12\sin\(\frac{\pi}{4}\text{ad}_P\) 
\end{align}
which we have just shown can be expressed as a weighted sum of the results obtained by sampling four different valid quantum channels that can each be implemented holographically.

This method becomes useful in cases where mid-circuit measurement errors are the dominant source of error (as opposed to, say, gate errors building up in the implementation of the qMPS tensors), and becomes helpful when the JW string is sufficiently long that its measurement error exceeds that introduced by the extra circuitry required to perform the $M_P$ and $\sin (\frac\pi4 \text{ad}_P)$ operations.

%\section{Additional data for GMPS+U scaling}

%\begin{figure}[t]
%\centering
%  \includegraphics[width=0.5\textwidth]{figures/GMPS+U_errvsL.pdf}
%    \caption{{\bf Error scaling versus system size --} the relative error in energy versus the total length $L$ of spinful sites for FH model at $U=4$ and $\nu = \frac{1}{2}$ and fixed block size $B=6$ 
%    \label{fig:ErrvsL} 
%    }
%\end{figure}

\section{Additional data for brick qMPS}
\label{appendix: additional}
% \begin{figure}[t] 
%     \centering
%     \includegraphics[width=0.5\textwidth]{figures/Fig2_VQE.pdf}
%     \caption{ {\bf Variational qMPS} (a) Unit cell and generating circuit structure of (infinite) variational MPS used for Fermi-Hubbard model holoVQE simulations. Directed lines indicate qubit world-lines. (b) Colored circles are number-conserving two-qubit gates, with $5$ variational parameters (chosen independently for each gate in the unit cell) of the form shown in (b), where $P_{\theta} = e^{-i\theta P/2}$ for any Pauli operator $P=\{X,Y,Z,XX,\dots\}$. 
%     \label{fig:vqe} 
%         }
% \end{figure}
In this section, we present additional simulation and experimental data for the problem-agnostic brick circuit qMPS for the Fermi-Hubbard chain at half-filling and \sfrac13-filling.
Noiseless simulations of the problem-agnostic brick-circuit qMPS approach show that it can effectively capture the ground-state and correlations over a range of $U$ and filling factors. 
To minimize the impact of errors, and reduce the implementation time for experimental demonstrations of this brick circuit qMPS, we first classically optimize the circuit parameters to minimize the variational energy for an infinite MPS (iMPS). From this, we classically compute the steady-state of the bond-transfer matrix and synthesize a circuit acting on the bond-qubits and one ancilla which approximately prepares this steady-state. This circuit is then used to prepare an initial mixed state of the bond qubits which closely approximates their bulk steady state, allowing us to directly access the infinite system-size limit without iteratively ``burning in'' the bond-channel as previously done in~\cite{foss2021holographic}. We note that a similar technique was employed by~\cite{barratt2021parallel}. We emphasize that this technique is only viable for small problem sizes where classical simulations are tractable. This, however, may still be useful for preparing, say, a moderate bond-dimension approximation of a correlated ground-state which is subsequently subjected to rapidly-entangling time evolution that could not be simulated classically.

The experimental data agrees well at larger values of $U$ but deviates significantly at $U=1$. Since the charge correlation length increases with smaller $U$'s, we interpret these deviations as arising from increased propagation of noise and errors in large-correlation length qMPS.

\bibliography{qmpsbib.bib}

\end{document}